\documentclass[10pt,onecolumn,twoside]{IEEEtran}
\usepackage{calc}

\usepackage{multirow}
\usepackage{cite}
\usepackage{graphicx,subfigure}
\usepackage{psfrag}
\usepackage{amsmath,amssymb}
\usepackage{color}
\usepackage{tikz}

\DeclareMathOperator*{\dotleq}{\overset{.}{\leq}}
\DeclareMathOperator*{\dotgeq}{\overset{.}{\geq}}
\DeclareMathOperator*{\defeq}{\triangleq}

\newtheorem{lemma}{Lemma}

\newtheorem{proposition}{Proposition}
\newtheorem{example}{Example}

\newcommand{\bit}{\begin{itemize}}
\newcommand{\eit}{\end{itemize}}

\newcommand{\bc}{\begin{center}}
\newcommand{\ec}{\end{center}}

\newcommand{\ba}{\begin{array}}
\newcommand{\ea}{\end{array}}

\newcommand{\beq}{\begin{equation}}
\newcommand{\eeq}{\end{equation}}

\newcommand{\beqn}{\begin{equation*}}
\newcommand{\eeqn}{\end{equation*}}

\newcommand{\bean}{\begin{eqnarray*}}
\newcommand{\eean}{\end{eqnarray*}}
\newcommand{\bea}{\begin{eqnarray}}
\newcommand{\eea}{\end{eqnarray}}

\def\E{\mathbb{E}}

\def\ev{\boldsymbol{e}}

\def\gv{\boldsymbol{g}}
\def\hv{\boldsymbol{h}}

\def\xv{\boldsymbol{x}}

\newcommand{\T}{{\scriptscriptstyle\mathsf{T}}}
\renewcommand{\H}{{\scriptscriptstyle\mathsf{H}}}

\newtheorem{remark}{Remark}

\renewcommand{\Bmatrix}[1]{\begin{bmatrix}#1\end{bmatrix}}

\begin{document}
\sloppy

\title{On the Vector Broadcast Channel with Alternating CSIT: A Topological Perspective}
\author{Jinyuan Chen, Petros Elia and Syed Ali Jafar
\thanks{Jinyuan Chen is with Stanford University, CA 94305, USA (email: jinyuanc@stanford.edu).
Petros Elia is with EURECOM, Sophia Antipolis, 06410, France (email: elia@eurecom.fr).
Syed Ali Jafar is with University of California, Irvine, CA 92697, USA (email: syed@uci.edu). 
The work of Jinyuan Chen was supported by the NSF Center for Science of Information: NSF-CCF-0939370. The work of Syed Ali Jafar was supported by NSF grants
CCF 1319104 and CCF 1317351. The work of Petros Elia was supported by the European Research Council under the European Community's Seventh Framework Programme (FP7/2007-2013) / grant agreement no. 257616 (CONECT), the FP7 CELTIC SPECTRA project, and  the Agence Nationale de la Recherche project ANR-IMAGENET.
}
\thanks{A smaller version of this paper will be presented at ISIT 2014.}
}


\maketitle
\thispagestyle{empty}

\begin{abstract}
In many wireless networks, link strengths are affected by many topological factors such as different distances, shadowing and inter-cell interference, thus resulting in some links being generally stronger than other links. From an information theoretic point of view, accounting for such topological aspects has remained largely unexplored, despite strong indications that such aspects can crucially affect transceiver and feedback design, as well as the overall performance.

The work here takes a step in exploring this interplay between topology, feedback and performance. This is done for the two user broadcast channel with random fading, in the presence of a simple two-state topological setting of statistically strong vs. weaker links, and in the presence of a practical ternary feedback setting of \emph{alternating channel state information at the transmitter} (alternating CSIT) where for each channel realization, this CSIT can be perfect, delayed, or not available.

In this setting, the work derives generalized degrees-of-freedom bounds and exact expressions, that capture performance as a function of feedback statistics and topology statistics. The results are based on novel \emph{topological signal management} (TSM) schemes that account for topology in order to fully utilize feedback. This is achieved for different classes of feedback mechanisms of practical importance, from which we identify specific feedback mechanisms that are best suited for different topologies. This approach offers further insight on how to split the effort --- of channel learning and feeding back CSIT --- for the strong versus for the weaker link. Further intuition is provided on the possible gains from topological spatio-temporal diversity, where topology changes in time and across users.
\end{abstract}

\section{Introduction}
A vector Gaussian broadcast channel, also known as the Gaussian MISO BC (multiple-input single-output broadcast channel) is comprised of a transmitter with multiple antennas that wishes to send independent messages to different receivers, each equipped with a single antenna. In addition to its direct relevance to cellular downlink communications, the MISO BC has attracted much attention for the critical role played in this setting by the feedback mechanism through which channel state information at the transmitter (CSIT) is typically acquired. Interesting insights into the dependence of the capacity limits of the MISO BC on the timeliness and quality of feedback, have been found through degrees of freedom (DoF) characterizations under perfect CSIT \cite{CS:03}, no CSIT \cite{JG:05,HJSV:12,LSW:05,VV:09}, compound CSIT \cite{WSK:07,GJW:11,MaddahAli:09}, delayed CSIT \cite{MAT:11c}, CSIT comprised of channel coherence patterns \cite{Jafar:12}, mixed CSIT \cite{YKGY:12d,GJ:12o,CE:12d,CE:13it}, and alternating CSIT \cite{TJSP:12}.
Other related work can be found in~\cite{GMK:11o,XAJ:11b,LSW:12,CE:13isit,CE:13spawc,LH:12,HC:13, CYE:13isit,CE:13MIMO,YYGK:12,VV:11t,VMA:13,GNS:07}.

\begin{figure}
\centering
\includegraphics[width=11cm]{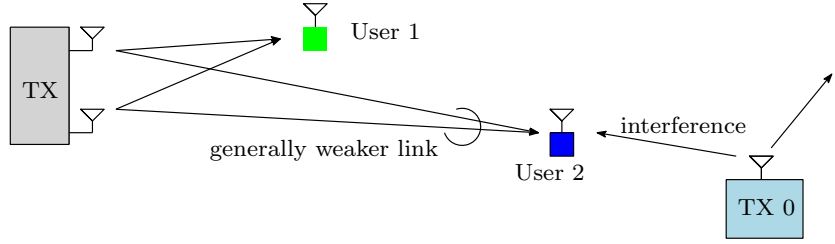}
\caption{Topology where link~2 is weaker due to distance and interference.}
\label{fig:DistanceChannel}
\end{figure}

\begin{figure}
\centering
\includegraphics[width=9cm]{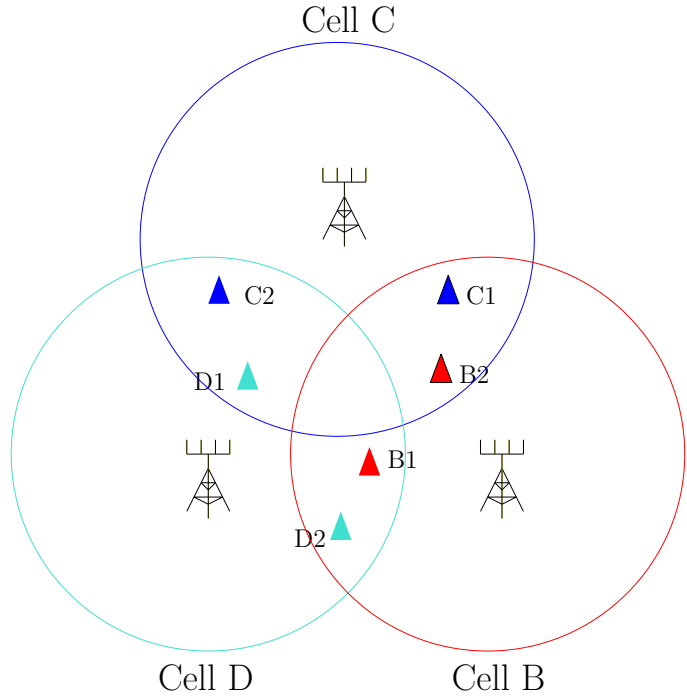}
\caption{Cell edge users experience fluctuating interference due to changing frequency allocation in the multi-cell system.}
\label{fig:Threecells}
\end{figure}

As highlighted recently in \cite{jafar:13}, while the insights obtained from DoF studies are quite profound, they are implicitly limited to settings where all users experience comparable signal strengths. This is due to the fundamental limitation of the DoF metric which treats each user with a non-zero channel coefficient, as capable of carrying exactly 1 DoF by itself, regardless of the statistical strength of the channel coefficients. Thus, the DoF metric ignores the diversity of link strengths, which is perhaps the most essential aspect of wireless communications from the perspective of interference management. Indeed, in wireless communication settings, the link strengths are affected by many topological factors, such as propagation path loss, shadow fading and inter-cell interference \cite{TV:05}, which lead to statistically unequal channel gains, with some links being much weaker or stronger than others (See Figures~\ref{fig:DistanceChannel}, \ref{fig:Threecells}).  Accounting for these topological aspects, by going beyond the DoF framework into the \emph{generalized} degrees of freedom (GDoF) framework, is the focus of the topological perspective that we seek here.

The work here combines considerations of topology with considerations of feedback timeliness and quality, and addresses questions on performance bounds, on encoding designs that account for topology and feedback, on feedback and channel learning mechanisms that adapt to topology, and on handling and even exploiting fluctuations in topology.

\section{System model for the topological BC \label{sec:system}}

\subsection{Channel, topology, and feedback models}
We consider the broadcast channel, with a two-antenna transmitter sending information to two single-antenna receivers.
The corresponding received signals at the first and second receiver at time $t$, can be modeled as
\begin{align}
y_t &=  \sqrt{\rho}\hv^{'\T}_t \xv_t  +  u^{'}_t      \label{eq:channely1}\\
z_t &=  \sqrt{\rho}\gv^{'\T}_t \xv_t  +  v^{'}_t      \label{eq:channely2}
\end{align}
where $\rho$ is defined by a power constraint, where $\xv_{t}$ is the normalized transmitted vector at time $t$ --- normalized here to satisfy $||\xv_{t}||^2  \le 1$ --- where $\hv^{'}_t,\gv^{'}_t$ represent the vector fading channels to the first and second receiver respectively, and where $u^{'}_t,v^{'}_t$ represent equivalent receiver noise.
\subsubsection{Topological diversity}
In the general topological broadcast channel setting, the variance of the above fading and equivalent noise, may be uneven across users, and may indeed fluctuate in time and frequency.  These fluctuations may be a result of movement, but perhaps more importantly, topological changes in the time scales of interest, can be attributed to fluctuating inter-cell interference. Such fluctuations are in turn due to different allocations of carriers in different cells or --- similarly --- due to the fact that one carrier can experience more interference from adjacent cells than another.

The above considerations can be concisely captured by the following simple model
\begin{align}
y_t &=  \rho^{A_{1,t}/2} \hv^{\T}_t \xv_t  +  u_t      \label{eq:channely21}\\
z_t &=  \rho^{A_{2,t}/2} \gv^{\T}_t \xv_t  +  v_t      \label{eq:channely22}
\end{align}
where now $\hv_t,\gv_t$ and $u_t,v_t$ are assumed to be spatially and temporally i.i.d\footnote{This suggests the simplifying formulation of unit coherence time.} Gaussian with zero mean and \emph{unit variance}. With $ ||\xv_{t}||^2 \le 1$, the parameter $\rho$ and the \emph{link power exponents} $A_{1,t},A_{2,t}$ reflect --- for each link, at time $t$ --- an \emph{average} received signal-to-noise ratio (SNR)
\begin{align}\label{eq:linkpower}
\mathbb{E}_{\hv_t,\xv_t} |\rho^{A_{1,t}/2}\hv^{\T}_t\xv_t|^2 & = \rho^{A_{1,t}}\\
\mathbb{E}_{\gv_t,\xv_t} |\rho^{A_{2,t}/2}\gv^{\T}_t\xv_t|^2 & = \rho^{A_{2,t}}.
\end{align}
In this setting we adopt a simple two-state topological model where the link exponents can each take, at a given time $t$, one of two values
\vspace{-7pt} \bc $A_{k,t}  \in \{ 1, \alpha\}  \quad \text{for}  \quad 0 \leq  \alpha \leq 1, \ \ k=1,2 $\ec
reflecting the possibility of either a strong link ($A_{k,t} =1$), or a weaker link ($A_{k,t} =\alpha$).
The adopted small number of topological states, as opposed to a continuous range of $A_{k,t}$ values, is motivated by static multi-carrier settings with adjacent cell interference, where the number of topological states can be proportional to the number of carriers.
\begin{remark}
We clarify that the rate of change of the topology --- despite the use of a common time index for $A_{k,t}$ and $\hv_t,\gv_t$ --- need not match in any way, the rate of change of fading.
We also clarify that our use of the term `link' carries a statistical connotation, so for example when we say that at time $t$ the first link is stronger than the second link, we refer to a statistical comparison where $A_{1,t}>A_{2,t}$.
\end{remark}

\subsubsection{Alternating CSIT formulation}
In terms of feedback, we draw from the alternating CSIT formulation by Tandon et al. \cite{TJSP:12}, which can nicely capture simple feedback policies. In this setting, the CSIT for each channel realization can be immediately available and perfect ($P$), or it can be delayed ($D$), or not available ($N$). In our notation, $I_{k,t}\in \{P,D,N\}$ will characterize the CSIT about the fading channel of user $k$ at time $t$.

\subsection{Problem statement: generalized degrees-of-freedom, feedback and topology statistics}

\subsubsection{Generalized Degrees-of-Freedom}
In a setting where $(R_1,R_2)$ denotes an achievable rate pair for the first and second user respectively, we focus on the high-SNR regime and seek to characterize sum GDoF
\[d_{\Sigma} = \lim_{\rho \to \infty} \max_{(R_1,R_2)}  \frac{R_1 +  R_2}{\log \rho}\]
performance bounds.

It is easy to see that in the current two-state topological setting, a strong link by itself has capacity that scales as $\log \rho + o (\log \rho)$, while\footnote{$o(\bullet)$ comes from the standard Landau notation, where $f(x) = o(g(x))$ implies $\lim_{x\to \infty} f(x)/g(x)=0$. Logarithms are of base~$2$.} a weak link has a capacity that scales as $\alpha\log \rho + o (\log \rho)$. Setting $\alpha = 1$ removes topology considerations, while setting $\alpha = 0$ almost entirely removes the weak link, as its capacity does not scale with SNR. Needless to say that setting the stronger link to correspond to a unit link-power exponent, is a result of normalization, and thus imposes no loss in generality.
\begin{example}
One can see that, in the current setting of the two-user MISO BC, having always perfect feedback ($P$) for both users' channels, and having a static topology where the first link is stronger than the second throughout the communication process ($A_{1,t}=1, A_{2,t}=\alpha, \ \forall t$), the sum GDoF is $d_\Sigma=1+\alpha$, and it is achieved by zero forcing.
\end{example}
\begin{example}\label{ex:MATorig}
Furthermore a quick back-of-the-envelope calculation (see the appendix in Section~\ref{sec:MATorig-DD1a}), can show that in the same fixed topology $A_{1,t}=1,A_{2,t}=\alpha, \forall t$, the original MAT scheme --- originally designed in \cite{MAT:11c} without topology considerations for the $\alpha = 1$ case --- after a small modification that regulates the rate of the private information to the weaker user, achieves a sum GDoF of $d_\Sigma=\frac{2}{3}(1 + \alpha)$. This performance will be surpassed by a more involved topological signal management (TSM) scheme, to be described later on.
\end{example}

\vspace{3pt}
\subsubsection{Motivation of the GDoF setting}
Often, taking a strict interpretation of the limiting nature of GDoF, leads to confusion because, strictly speaking, any reasonable channel model would force a limiting $\alpha$ to be 1, since all powers would go to infinity the same way. Towards convincing the skeptical reader of the usefulness of our approach, we offer the following thoughts which can help clarify any misconceptions.

Our GDoF approach here is based on two crucial premises.\\
$i)$ Network links generally have different capacities, and in the perfectly conceivable case where a link has a capacity that is a fraction $\alpha$ of another link's capacity, a good approximation is that the weaker link has average power that is close to the $\alpha^{th}$ power of the aforementioned power of the strong link.\\
$ii)$ Albeit depending on the \emph{limiting} behavior of random variables, our result here can also be interpreted in the \emph{large} SNR regime, where you pick $\alpha$ based on the aforementioned premise, and once this $\alpha$ is picked and fixed, the high-SNR approximation can yield expressions which, for sufficiently large SNR, have a gap from reality that is expected to be substantially smaller than the derived expression --- thus allowing for the derived expression to offer a good qualitative estimate of the overall behavior. Deviating from the strict and literal interpretation of GDoF, while still mathematically rigorous, the current approach allows us to consider topological settings that are motivated by reasonable scenarios that include distance variations and interference fluctuations, and does not constrain us to `limiting' awkward scenarios where variable geometries have distances that scale in different specific ways.

\subsubsection{Feedback and topology statistics}
Naturally performance is a function of the feedback and topology statistics. In terms of feedback statistics, we draw from the formulation in~\cite{TJSP:12} and consider
\vspace{-4pt}\bc $\lambda_{I_1,I_2}$\ec
to denote the fraction of the time during which the CSIT state is described by a pair $(I_1,I_2)\in (P,D,N) \times (P,D,N)$. 

We similarly consider
\vspace{-5pt}\bc $\lambda_{A_1,A_2}$\ec
to denote the fraction of the time during which the gain exponents of the two links are some pair $(A_1,A_2) \in (1,\alpha)\times (1,\alpha)$, where naturally $\lambda_{1,\alpha} + \lambda_{\alpha,1} + \lambda_{1,1} + \lambda_{\alpha,\alpha} = 1.$
Finally we use
\vspace{-4pt}\bc$\lambda_{I_1,I_2}^{A_1,A_2}$\ec to denote the fraction of the time during which the CSIT state is $(I_1,I_2)$ and the topology state is $(A_1,A_2)$.

\begin{example}
$\lambda_{P,P} = 1$ (resp. $\lambda_{D,D} = 1$, $\lambda_{N,N} = 1$) implies perfect CSIT (resp. delayed CSIT, no CSIT) for both users' channels, throughout the communication process. Similarly $\lambda_{P,N}+\lambda_{N,P} = 1$ restricts to a family of feedback schemes where only one user sends CSIT at a time (more precisely, per channel realization), and does so perfectly. From this family, $\lambda_{P,N}=\lambda_{N,P} = 1/2$ is the symmetric option.
Similarly, in terms of topology, $\lambda_{1,\alpha}=1, \ \alpha<1$ implies that the first link is stronger than the second throughout the communication process, while $\lambda_{1,\alpha}=\lambda_{\alpha,1}=1/2$ implies that half of the time, the first user is statistically stronger, and vice versa.\\
Finally having $\lambda_{P,D}^{1,\alpha}+\lambda_{D,P}^{\alpha,1} = 1$ does not impose any restriction on the topology statistics, but it implies a feedback mechanism that asks --- for any channel realization --- the statistically stronger user to send perfect feedback, and the statistically weaker user to send delayed feedback.
\end{example}

\subsection{Conventions and structure}
In terms of notation, $(\bullet)^\T$ and $(\bullet)^{\H}$ denote the transpose and conjugate transpose operations respectively, while $||\bullet||$ denotes the Euclidean norm, and $|\bullet|$ denotes either the magnitude of a scalar or the cardinality of a set.
We also use $\doteq$ to denote \emph{exponential equality}, i.e., we write $f(\rho)\doteq \rho^{B}$ to denote $\displaystyle\lim_{\rho\to\infty}\log f(\rho)/\log \rho=B$. Similarly $\dotgeq$ and $\dotleq$ denote exponential inequalities.
$\ev^{\bot}$ denotes a unit-norm vector orthogonal to vector $\ev$.

Throughout this paper, we adhere to the common convention and assume perfect and global knowledge of channel state information at the receivers (perfect and global CSIR).

We proceed with the main results.  We first present sum GDoF outer bounds as a function of the CSIT and topology statistics, and then proceed to derive achievable and often optimal sum GDoF expressions for pertinent cases of practical significance.

\section{Outer bounds}

We first proceed with a simpler version of the outer bound, which encompasses all cases of alternating CSIT, and all \emph{fixed} topologies ($\lambda_{1,\alpha} = 1$, or $\lambda_{\alpha,1} = 1, \ \alpha\in[0,1]$).
\vspace{3pt}
\begin{lemma}\label{lem:outerbSImple}
The sum GDoF of the two-user MISO BC with alternating CSIT and a fixed topology, is upper bounded as $d_\Sigma  \leq   \min\{d_{\Sigma}^{(1)}, d_{\Sigma}^{(2)}\}$, where
\begin{align*}
d_{\Sigma}^{(1)} & \defeq   (1+\alpha)  \lambda_{ P,P}   +  \frac{3 + 2 \alpha}{3}  (  \lambda_{ P,D} \!+\! \lambda_{ D,P} \!+\!  \lambda_{ P,N}  \!+\!  \lambda_{N,P}  )  \nonumber \\  &\quad +  \frac{3 +  \alpha}{3}  ( \lambda_{ D,D}      +   \lambda_{ D,N}  +  \lambda_{ N,D}  +  \lambda_{ N,N} )
\nonumber \\
d_{\Sigma}^{(2)} & \defeq      (1+ \alpha )( \lambda_{ P,P} +  \lambda_{ P,D} +\lambda_{ D,P} +\lambda_{ D,D} )  \nonumber \\  &\quad   + \frac{2+ \alpha }{2}(\lambda_{ P,N} +\lambda_{ N,P} + \lambda_{ D,N} +\lambda_{ N,D} )   +   \lambda_{ N,N}.
\end{align*}
\end{lemma}
\vspace{3pt}
The proof of the above lemma, can be found as part of the proof of the following more general lemma, in the appendix of Section~\ref{sec:outerbProof}.

We now proceed with the general outer bound, for any alternating CSIT mechanism, and any topology, i.e., for any $\lambda_{I_1,I_2}^{A_1,A_2}$.
For conciseness we use
\begin{align}
\lambda_{ P \leftrightarrow N}^{ A_1, A_2} &\defeq  \lambda_{ P,N}^{A_1, A_2}+ \lambda_{N,P}^{A_1, A_2} \nonumber\\
\lambda_{ D \leftrightarrow N}^{ A_1, A_2} &\defeq  \lambda_{ D,N}^{A_1, A_2}+ \lambda_{N,D}^{A_1, A_2} \nonumber\\
\lambda_{ P \leftrightarrow D}^{ A_1, A_2} &\defeq   \lambda_{ P,D}^{A_1, A_2}+ \lambda_{D,P}^{A_1,A_2}\nonumber
\end{align}
so for example, $ \lambda_{P\leftrightarrow D}^{1, \alpha}$ simply denotes the fraction of the communication time during which the first link is stronger than the second, and during which, the CSIT for the channel of \emph{any one} of the users, is being fed back in a perfect and instantaneous manner, while the CSIT for the channel of the other user, is fed back later in a delayed manner.

\vspace{3pt}
\begin{lemma} \label{lem:outerb}
The sum GDoF of the topological two-user MISO BC with alternating CSIT, is upper bounded as
\beq
d_\Sigma \leq   \min\{d_{\Sigma}^{(3)}, d_{\Sigma}^{(4)}\}    \label{eq:outerbound}
\eeq
where
\begin{align}
d_{\Sigma}^{(3)} &\defeq   (1+\alpha)  (  \lambda_{P,P}^{ \alpha, 1}  +  \lambda_{ P,P}^{ 1, \alpha}   )  +  \frac{3 + 2 \alpha}{3}  (  \lambda_{ P\leftrightarrow D}^{\alpha, 1}  +  \lambda_{ P\leftrightarrow D}^{1, \alpha}   )  +  \frac{3 + 2 \alpha}{3}  (  \lambda_{P\leftrightarrow N}^{\alpha, 1}  +  \lambda_{P\leftrightarrow N}^{1, \alpha}   )
\nonumber\\  & \quad  +  \frac{3 +  \alpha}{3}  (  \lambda_{D,D}^{\alpha, 1}  +  \lambda_{D,D}^{1, \alpha}   )   +  \frac{3 +  \alpha}{3}  (  \lambda_{D\leftrightarrow N}^{\alpha, 1}  +  \lambda_{D\leftrightarrow N}^{1, \alpha}   )   +  \frac{3 +  \alpha}{3}  (  \lambda_{N,N}^{\alpha, 1}  +  \lambda_{N,N}^{1, \alpha}   )   \nonumber\\  & \quad  +  2  \lambda_{P,P}^{ 1, 1}   + \frac{ 5 }{3}  \lambda_{P\leftrightarrow D}^{ 1, 1} + \frac{ 5 }{3}  \lambda_{P\leftrightarrow N}^{ 1, 1} + \frac{ 4 }{3}  \lambda_{D,D}^{ 1, 1}  + \frac{ 4 }{3}   \lambda_{D\leftrightarrow N}^{1, 1} + \frac{ 4 }{3}   \lambda_{N,N}^{ 1, 1}
\nonumber\\  & \quad   +  2\alpha  \lambda_{P,P}^{ \alpha, \alpha}
  + \frac{ 5\alpha }{3} \lambda_{P\leftrightarrow D}^{ \alpha, \alpha}
    + \frac{ 5\alpha }{3}  \lambda_{P\leftrightarrow N}^{ \alpha, \alpha}
     +  \frac{4\alpha }{3}  \lambda_{D,D}^{ \alpha, \alpha}
   + \frac{4\alpha }{3} \lambda_{D\leftrightarrow N}^{\alpha, \alpha}
   + \frac{4\alpha }{3}  \lambda_{N,N}^{ \alpha, \alpha}   \label{eq:outerbound1} \\
d_{\Sigma}^{(4)}  &\defeq
(1+ \alpha )(\lambda_{P,P}^{ 1, \alpha} +\lambda_{P,P}^{ \alpha, 1} )  +   (1+ \alpha )(\lambda_{P\leftrightarrow D}^{ 1, \alpha} +\lambda_{P\leftrightarrow D}^{ \alpha, 1} )
+   (1+ \alpha )(\lambda_{D,D}^{ 1, \alpha} +\lambda_{D,D}^{ \alpha, 1} )
\nonumber\\  & \quad + \frac{2+ \alpha }{2}(\lambda_{P\leftrightarrow N}^{ 1, \alpha} +\lambda_{P\leftrightarrow N}^{ \alpha, 1} )   + \frac{2+ \alpha }{2}(\lambda_{D\leftrightarrow N}^{ 1, \alpha} +\lambda_{D\leftrightarrow N}^{ \alpha, 1} )   +   \lambda_{N,N}^{ 1, \alpha} +\lambda_{N,N}^{ \alpha, 1}   \nonumber\\  & \quad  +  2 \lambda_{P,P}^{ 1, 1} + 2\alpha \lambda_{P,P}^{ \alpha, \alpha} +  2 \lambda_{ P\leftrightarrow D}^{ 1, 1} + 2\alpha \lambda_{P\leftrightarrow D}^{ \alpha, \alpha}
+ 2 \lambda_{ D,D}^{ 1, 1} + 2\alpha \lambda_{D,D}^{ \alpha, \alpha}
\nonumber\\  & \quad  +   \frac{3}{2} \lambda_{ P\leftrightarrow N}^{ 1, 1} + \frac{3\alpha}{2} \lambda_{ P\leftrightarrow N}^{ \alpha, \alpha}
+   \frac{3}{2} \lambda_{ D\leftrightarrow N}^{ 1, 1} + \frac{3\alpha}{2} \lambda_{ D\leftrightarrow N}^{ \alpha, \alpha}       +    \lambda_{ N,N}^{ 1, 1} +\alpha\lambda_{ N,N}^{ \alpha, \alpha}.   \label{eq:outerbound2}
\end{align}
\end{lemma}

The above bounds will be used to establish the optimality of different encoding schemes and practical feedback mechanisms.

\section{Practical feedback schemes over a fixed topology}

We first proceed to derive different results for the case of any \emph{fixed} topology. Here, without loss of generality, we will consider the case where $\lambda_{1,\alpha} = 1$, while the case of $\lambda_{\alpha,1} = 1$ is handled simply by interchanging the role of the two users. In the presence of a fixed topology, we initially focus on different practical feedback schemes for which we derive the exact sum GDoF expressions, and then proceed to explore the delayed CSIT case for which we derive a bound.

With emphasis on practicality, we first focus on three families of simple feedback mechanisms which can be implemented so that, per coherence interval, only one user sends feedback\footnote{In our formulation, which uses the simplifying assumption of having a unit coherence period, this simply refers to the case where only one user sends feedback at a time.}.

\vspace{3pt}
\begin{proposition} \label{prop:simpleFB-fixedTopology}
For the two-user MISO BC with a fixed topology and a feedback constraint $\lambda_{P,N}+\lambda_{N,P} = 1$ or $\lambda_{P,N}+\lambda_{N,P}=\lambda_{N,D}+\lambda_{D,N} = 1/2$ or $\lambda_{P,D}+\lambda_{D,P}=\lambda_{N,N} = 1/2$, the optimal sum GDoF is
\begin{align}
d_{\Sigma} = 1+  \frac{ \alpha}{2}
\end{align}
where in the first case, this is achieved by the symmetric mechanism $\lambda_{P,N}= \lambda_{N,P}= 1/2$, in the second case it is achieved by the symmetric mechanism $\lambda_{P,N}= \lambda_{N,D}= 1/2$ which associates delayed feedback with the weak user, and in the third case it is achieved by the mechanism $\lambda_{P,D}= \lambda_{N,N}= 1/2$, which again associates delayed feedback with the weak user.
\end{proposition}
\vspace{3pt}
\begin{proof}
All GDoF expressions are optimal as they meet the outer bound in Lemma~\ref{lem:outerbSImple}. For the first case ($\lambda_{P,N}+\lambda_{N,P} = 1$) the GDoF optimal scheme can be found in Section~\ref{sec:PN1a-NP1a}, for the case where $\lambda_{P,N}+\lambda_{N,P}=\lambda_{N,D}+\lambda_{D,N} = 1/2$ the optimal scheme can be found in Section~\ref{sec:ND1a-PN1a}, while for the last case where $\lambda_{P,D}+\lambda_{D,P}=\lambda_{N,N} = 1/2$ the optimal scheme can be found in Section~\ref{sec:PD1aNN1a}.
\end{proof}

\begin{remark}
The optimality of $\lambda_{P,N}= \lambda_{N,D}= 1/2$ (resp. $\lambda_{P,D}= \lambda_{N,N}= 1/2$) among all possible mechanisms $\lambda_{P,N}+\lambda_{N,P} = \lambda_{D,N}+\lambda_{N,D} = 1/2$ (resp. $\lambda_{P,D}+\lambda_{D,P}=\lambda_{N,N} = 1/2$), relates to the fact that delayed CSIT is associated to the weak link, which in turn allows for the unintended interference --- resulting from communicating without current CSIT --- to be naturally reduced in the direction of the weak link.
\end{remark}

\begin{remark}
It is easy to see that the family $\lambda_{P,D}+\lambda_{D,P}=\lambda_{N,N} = 1/2$ is again a `one-user-per-channel'  family of feedback policies since it can be implemented by having half of the channel states not fed back, while having the other half fed back by any one user with no delay, and by the other user with delay.
\end{remark}
\vspace{3pt}

\subsection{Delayed CSIT and fixed topology}
For the same setting of fixed topologies ($\lambda_{1, \alpha}=1$ or $\lambda_{\alpha,1}=1$, $\alpha\in[0,1]$), we lower bound the sum GDoF performance for the well known delayed CSIT scenario of Maddah-Ali and Tse \cite{MAT:11c}, where feedback is always delayed ($\lambda_{D,D}=1$). 
\vspace{3pt}
\begin{proposition} \label{prop:DD1a}
For the two-user MISO BC with a fixed topology and delayed CSIT ($\lambda_{D,D}=1$), the sum GDoF is lower bounded as
\begin{align}
d_\Sigma \geq 1 + \frac{\alpha^2}{2+\alpha}.
\end{align}
\end{proposition}
\vspace{3pt}
\begin{proof}
The scheme that achieves the lower bound can be found in Section~\ref{sec:DD1a}. 
\end{proof}

It is worth noting that the above sum GDoF surpasses the aforementioned performance of the original --- and slightly modified MAT scheme \cite{MAT:11c} --- over the same topology, which was mentioned in example~\ref{ex:MATorig} to be $d_\Sigma = \frac{2}{3}(1 + \alpha)$.

\section{Optimal sum GDoF of practical feedback schemes for the BC with topological diversity}
We here explore a class of alternating topologies and reveal a gain --- in certain instances --- that is associated to topologies that vary in time and across users. Emphasis is mainly given to statistically symmetric topologies.

We first proceed, and for the delayed CSIT setting $\lambda_{D,D}=1$, derive the optimal sum GDoF in the presence of the symmetrically \emph{alternating topology} where $\lambda_{1,\alpha}=\lambda_{\alpha,1}=1/2$.

\vspace{3pt}
\begin{proposition} \label{prop:DD1a-DDa1}
For the two-user MISO BC with delayed CSIT $\lambda_{D,D}=1$ and topological spatio-temporal diversity such that $\lambda_{1, \alpha} = \lambda_{\alpha, 1}=1/2$, the optimal sum GDoF is
\begin{align}
d_{\Sigma} = 1+  \frac{\alpha}{3}
\end{align}
which can be seen to exceed the optimal sum GDoF $d^{'}_\Sigma = \frac{2}{3}(1 +  \alpha) $ of the same feedback scheme, over an equivalent\footnote{The compared topologies are considered equivalent in the sense that the overall duration of weak links, is the same for the two topologies.} but spatially non-diverse topology $\lambda_{1, 1} =\lambda_{\alpha, \alpha}=1/2 $.
\end{proposition}
\vspace{3pt}
\begin{proof}
The GDoF is optimal as it meets the general outer bound in Lemma~\ref{lem:outerb}. The optimal TSM scheme is described in Section~\ref{sec:DD1a-DDa1}.
\end{proof}

\begin{figure}
\centering
\includegraphics[width=10cm]{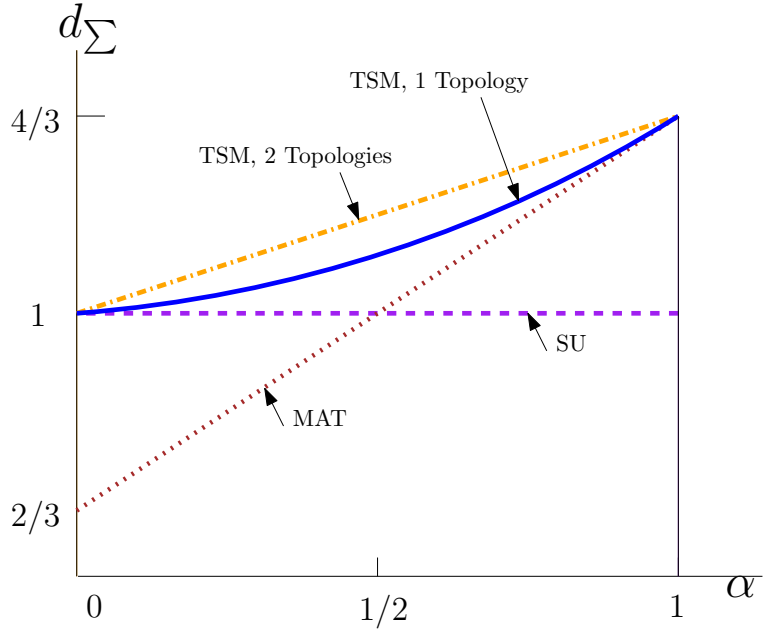}
\caption{Sum GDoF performance for the slightly modified Maddah-Ali and Tse scheme (MAT), the single user case (SU), and the topological signal management scheme (TSM 1), all for the setting $\lambda_{D,D}^{1,\alpha}=1$. Additionally the plot (TSM 2) describes the optimal sum GDoF for the fluctuating topology setting where $\lambda_{D,D}^{1, \alpha} = \lambda_{D,D}^{\alpha, 1}=1/2$. }
\label{fig:DCSITresults}
\end{figure}

We also briefly note that for the same feedback policy $ \lambda_{D,D}=1$, the optimal sum GDoF $d_{\Sigma} = 1+  \frac{\alpha}{3}$ corresponding to the topologically diverse setting $\lambda_{1, \alpha} = \lambda_{\alpha, 1}=1/2$, also exceeds the sum GDoF performance in Proposition~\ref{prop:DD1a} of the TSM scheme in the presence of any static topology (e.g. $\lambda_{1, \alpha} = 1$).

A similar observation to that of the above proposition, is derived below, now for the feedback mechanism $ \lambda_{P,N}=\lambda_{N,P}=1/2$.
\begin{proposition} \label{prop:PNgeneral-altTopology}
For the two-user MISO BC with $ \lambda_{P,N}=\lambda_{N,P}=1/2$ and topological diversity such that $\lambda_{1, \alpha} = \lambda_{\alpha, 1}=1/2$, the optimal sum GDoF is
\begin{align}
d_{\Sigma} = 1+  \frac{\alpha}{2}
\end{align}
which can be seen to exceed the optimal sum GDoF $d^{'}_\Sigma = \frac{3}{4}(1 +  \alpha) $ of the same feedback mechanism over the equivalent but spatially non-diverse topology $ \lambda_{1, 1} =\lambda_{\alpha, \alpha}=1/2 $.
\end{proposition}
\vspace{3pt}
\begin{proof}
The sum GDoF is optimal as it achieves the general outer in Lemma~\ref{lem:outerb}. The optimal scheme is described in Section~\ref{sec:PNNP-any}.
\end{proof}

Regarding this same feedback policy $\lambda_{P,N}= \lambda_{N,P}= 1/2$, it is worth to now note this policy's very broad applicability. This is shown in the following proposition.

\vspace{3pt}
\begin{proposition} \label{prop:simpleFB-2Topologies}
For the two-user MISO BC with any strictly uneven topology $\lambda_{1, \alpha}  + \lambda_{\alpha,1} =1$ and a feedback constraint $\lambda_{P,N}+\lambda_{N,P} = 1$, the optimal sum GDoF is
\begin{align}
d_{\Sigma} = 1+  \frac{ \alpha}{2}
\end{align}
and it is achieved by the symmetric feedback policy $\lambda_{P,N}= \lambda_{N,P}= 1/2$.
\end{proposition}
\vspace{3pt}
\begin{proof}
The sum GDoF is optimal as it achieves the general outer bound in Lemma~\ref{lem:outerbSImple}.
The optimal scheme is described in Section~\ref{sec:PNNP-any}.
\end{proof}

\begin{remark}
This broad applicability of mechanism $\lambda_{P,N}= \lambda_{N,P}= 1/2$, implies a simpler process of learning the channel and generating CSIT, which now need not consider the specific topology as long as this is strictly uneven ($\lambda_{1,1} =\lambda_{\alpha,\alpha} =0$).
\end{remark}

\section{Conclusions\label{sec:conclusion}}

The work explored the interplay between topology, feedback and performance, for the specific setting of the two-user MISO broadcast channel. Adopting a generalized degrees of freedom framework, and addressing feedback and topology jointly, the work revealed new aspects on encoding design that accounts for topology and feedback, as well as new aspects on how to handle and even exploit topologically diverse settings where the topology varies across users and across time.

In addition to the bounds and encoding schemes, the work offers insight on how to feedback --- and naturally how to learn  --- the channel in the presence of uneven and possibly fluctuating topologies. This insight came in the form of simple feedback mechanisms that achieve optimality --- under specific constraints --- often without knowledge of topology and its fluctuations.

\section{Appendix - Proof of general outer bound (Lemma~\ref{lem:outerb}) \label{sec:outerbProof}}
We here provide the proof of the general outer bound in Lemma~\ref{lem:outerb}.
Let $W_1,W_2$ respectively denote the messages of user~1 and user~2, let $R_1,R_2$ denote the two users' rates, and let $\Omega^{n}$ denote all channel states that appear in the BC.  Let the communication duration be $n$ channel uses, where $n$ is large.
We use $$y^{n}_{I_1,I_2} = \{y_t\}_{t}, \ \ \ \ \ z^{n}_{I_1,I_2} = \{z_t\}_{t} \ \ \ \ \forall t: I_{1,t} = I_1,I_{2,t} = I_2$$ to denote the accumulated set of received signals at user~1 and user~2 respectively, accumulated throughout the time when the CSIT state was some fixed $I_1,I_2$.
As a result, the entirety of the received signals, at each user, is the union of the above sets
\[    y^{n} = \bigcup_{I_1,I_2} y^{n}_{I_1,I_2} ,   \ \ \ \ \ \ \ z^{n} = \bigcup_{I_1,I_2} z^{n}_{I_1,I_2} .  \]

\subsection{Proof for $d_1+ d_2\leq d_{\sum}^{(3)}$  \ (cf. \eqref{eq:outerbound1}) }

We first enhance the BC by offering user~2, complete knowledge of $y^{n}$ and of $W_1$.
Having now constructed a degraded BC, we proceed to remove all delayed feedback. This removal, which is equivalent to substituting the CSIT state $I_k = D$ with $I_k = N$, does not affect capacity, as one can deduce from the work in~\cite{Gamal:78t}.

We then proceed to construct a degraded compound BC by adding an additional user, denoted as user~$\tilde{1}$, seeking to receive the same desired message $W_1$ as user~1.
The received signal of user~$\tilde{1}$ takes the form
\[    \tilde{y}^{n} = \bigl( y^{n}_{P,P} ,  y^{n}_{P,D},  y^{n}_{P,N}, \tilde{y}^{n}_{D,P}, \tilde{y}^{n}_{N,P},   \tilde{y}^{n}_{D,D}, \tilde{y}^{n}_{D,N},  \tilde{y}^{n}_{N,D}, \tilde{y}^{n}_{N,N} \bigr)  \]
where specifically when $I_1 = P$ (i.e., whenever the first user sends perfect CSIT) then the received signal of user~$\tilde{1}$ is identical to that of user~1, else when $I_1 \neq P$, the received signal of user~$\tilde{1}$ is only assumed to be \emph{identically distributed} to the signal $y_t$ of user~1.
We also assume that throughout the communication process, user~$\tilde{1}$ and user~1 experience the same channel gain exponent $A_{1,t}$ for all $t$  (cf. \eqref{eq:channely21}).
We further enhance by assuming that $\tilde{y}^{n} $ is known to user~2.
We note that, since user~1 and user~$\tilde{1}$ have the same decodability, the capacity of this degraded compound BC cannot be worse than that of the original degraded BC.

As a next step, we introduce the auxiliary random variable $s_t$, and define $s^{n}_{I_1,I_2} = \{s_t\}_{t: I_{1,t}=I_1,I_{2,t}=I_2}$.
At this point we enhance the degraded compound BC, by giving user~2 complete knowledge of
\[s^{n}_{0}\defeq \{  s^{n}_{D,P},s^{n}_{N,P},s^{n}_{D,N},s^{n}_{N,D}, s^{n}_{D,D}, s^{n}_{N,N}\} \]
where, as described below in \eqref{eq:SsignalDef}, $\{s^{n}_{D,P},s^{n}_{N,P},s^{n}_{D,N},s^{n}_{N,D}, s^{n}_{D,D}, s^{n}_{N,N}\}$ is the collection of auxiliary random variables $s_t, \ t: I_{1,t}\neq P$ accumulated whenever there is no CSIT on channel $\hv_{t}$ of user~1 and no CSIT on channel $\tilde{\hv}_{t}$ of user~$\tilde{1}$, where specifically
\begin{align}  \label{eq:SsignalDef}
 \rho^{\frac{A_{2,t} - A_{1,t} }{2} } \Bmatrix{ \hv_t^\T \\  \gv_t^\T }    \Bmatrix{ \hv_{t}^\T \\ \tilde{\hv}_{t}^\T }^{-1}
\Bmatrix{y_{t} \\ \tilde{y}_{t}}     = \underbrace{ \rho^{\frac{ A_{2,t} }{2} }   \Bmatrix{ \hv_t^\T \\  \gv_t^\T }   \xv_t  +   \Bmatrix{ 0 \\  v_t }  }_{  = \Bmatrix{  \star  \\    z_t}    }
+   \underbrace{\Bmatrix{ 0 \\  - v_t }    +          \rho^{\frac{A_{2,t} - A_{1,t} }{2} } \Bmatrix{ \hv_t^\T \\  \gv_t^\T }    \Bmatrix{ \hv_{t}^\T \\ \tilde{\hv}_{t}^\T }^{-1}
\Bmatrix{  u_t   \\   \tilde{u}_{t}} }_{ \defeq \Bmatrix{  \star  \\    s_t}    }
\end{align}
i.e., where specifically $s_t$ is the second element of the vector $\Bmatrix{ 0 \\  - v_t }    +          \rho^{\frac{A_{2,t} - A_{1,t} }{2} } \Bmatrix{ \hv_t^\T \\  \gv_t^\T }    \Bmatrix{ \hv_{t}^\T \\ \tilde{\hv}_{t}^\T }^{-1}
\Bmatrix{  u_t   \\   \tilde{u}_{t}}$, and where we have set $\tilde{\hv}_{t}$ to be independently and identically distributed to $\hv_{t}$, and $\tilde{u}_{t}$ to be independently and identically distributed to $u_{t}$. What the above means is that $s_t$ has average power $$\E|s_t|^2 \doteq  \rho^{(A_{2,t} - A_{1,t}  )^{+}} $$ as well as that knowledge of  $\{s_t, y_{t}, \tilde{y}_{t}, \Omega^{n}\}$, implies the knowledge of  $z_{t}$, again whenever $I_1\neq P$.

At this point we can see that
\begin{align}
& nR_1-  n \epsilon_n \nonumber\\
& = H(W_1) - n \epsilon_n \nonumber\\
&=  H(W_1|\Omega^{n}) - n \epsilon_n  = I(W_1;y^{n}|\Omega^{n})+\underbrace{H(W_1|y^{n},\Omega^{n})}_{\leq n \epsilon_n}- n \epsilon_n \nonumber\\
&\leq I(W_1; y^{n}|\Omega^{n})  \label{eq:bound1R1fano}\\
&=  h(y^{n}|\Omega^{n})  - h(y^{n}|W_1, \Omega^{n}) \label{eq:bound1R1201}
\end{align}
where \eqref{eq:bound1R1fano} results from Fano's inequality which bounds $H(W_1|y^{n},\Omega^{n})$.

Similarly, for virtual user~$\tilde{1}$, we have
\begin{align}
& nR_1-  n \epsilon_n \nonumber\\
&\leq  h(\tilde{y}^{n}|\Omega^{n})  - h(\tilde{y}^{n}|W_1, \Omega^{n}).  \label{eq:bound1R1521}
\end{align}

As a result, adding \eqref{eq:bound1R1201} and \eqref{eq:bound1R1521} gives
\begin{align}
& 2 nR_1-  2n \epsilon_n \nonumber\\
&\leq  h(y^{n}|\Omega^{n}) +h(\tilde{y}^{n}|\Omega^{n})   - h(y^{n}|W_1, \Omega^{n})   - h(\tilde{y}^{n}|W_1, \Omega^{n}) \nonumber\\
&\leq  h(y^{n}|\Omega^{n}) +h(\tilde{y}^{n}|\Omega^{n})   - h(y^{n}, \tilde{y}^{n} |W_1, \Omega^{n}) \label{eq:2R1sum}
\end{align}
where \eqref{eq:2R1sum} uses a basic entropy inequality.

Now recalling that user~2 has knowledge of $\{W_1,z^{n},y^{n}, \tilde{y}^{n}, s^{n}_{0}\}$, gives
\begin{align}
& nR_2-  n \epsilon_n \nonumber\\
& = H(W_2) - n \epsilon_n \nonumber\\
&=  H(W_2|\Omega^{n}) - n \epsilon_n \nonumber\\
&\leq I(W_2; W_1,z^{n},y^{n}, \tilde{y}^{n}, s^{n}_{0}|\Omega^{n})  \label{eq:fano102}\\
&= I(W_2; z^{n},y^{n}, \tilde{y}^{n}, s^{n}_{0} |W_1,\Omega^{n})  + \underbrace{I(W_2; W_1|\Omega^{n})}_{  =  0} \\
&= h(z^{n},y^{n}, \tilde{y}^{n}, s^{n}_{0} |W_1,\Omega^{n}) -  \underbrace{h(z^{n},y^{n}, \tilde{y}^{n}, s^{n}_{0} |W_1,W_2,\Omega^{n})}_{  =  no(\log\rho) }    \\
&=  h(z^{n},y^{n}, \tilde{y}^{n}, s^{n}_{0} |W_1,\Omega^{n}) -    no(\log\rho)       \label{eq:entroRecon}\\
&=    h(y^{n}, \tilde{y}^{n} |W_1,\Omega^{n}) +\underbrace{ h(s^{n}_{0} |y^{n}, \tilde{y}^{n}, W_1,\Omega^{n})}_{\leq  h(s^{n}_{0})  }   +  h(z^{n}|y^{n}, \tilde{y}^{n}, s^{n}_{0}, W_1,\Omega^{n})   -    no(\log\rho)     \label{eq:chainrule}  \\
&\leq    h(y^{n}, \tilde{y}^{n} |W_1,\Omega^{n}) +  h(s^{n}_{0})     +  h(z^{n}|y^{n}, \tilde{y}^{n}, s^{n}_{0}, W_1,\Omega^{n})   -    no(\log\rho)  \label{eq:CondEntrop210}  \\
&\leq    h(y^{n}, \tilde{y}^{n} |W_1,\Omega^{n}) +  h(s^{n}_{0})     +  h(z^{n}_{P,P},z^{n}_{P,D},z^{n}_{P,N})  \nonumber\\ &\quad +  \underbrace{ h(z^{n}_{D,P},z^{n}_{N,P},z^{n}_{D,N},z^{n}_{N,D}, z^{n}_{D,D},z^{n}_{N,N}|y^{n}, \tilde{y}^{n}, s^{n}_{0}, W_1,\Omega^{n})}_{  \leq  no(\log\rho)  }  -    no(\log\rho)    \label{eq:CondEntrop834} \\
&\leq    h(y^{n}, \tilde{y}^{n} |W_1,\Omega^{n}) +  h(s^{n}_{0})     +  h(z^{n}_{P,P},z^{n}_{P,D},z^{n}_{P,N})  +  no(\log\rho)    \label{eq:ReconEntroAll}
\end{align}
where \eqref{eq:fano102} comes from Fano's inequality,
where \eqref{eq:entroRecon} follows from $h(z^{n},y^{n}, \tilde{y}^{n}, s^{n}_{0} |W_1,W_2,\Omega^{n}) = h(z^{n},y^{n}, \tilde{y}^{n} |W_1,W_2,\Omega^{n}) + \underbrace{h(s^{n}_{0} |z^{n},y^{n}, \tilde{y}^{n},W_1,W_2,\Omega^{n})}_{=0}  = h(z^{n},y^{n}, \tilde{y}^{n} |W_1,W_2,\Omega^{n})   =   no(\log\rho) $ by using the fact that the knowledge of $\{z^{n},y^{n}, \tilde{y}^{n},\Omega^{n}\}$ allows for the reconstruction of  $s^{n}_{0}$ (cf. \eqref{eq:SsignalDef}) and the fact that the knowledge of  $\{W_1,W_2,\Omega^{n}\}$ allows for reconstructing  $\{z^{n},y^{n}, \tilde{y}^{n}\}$ up to noise level, 
where \eqref{eq:chainrule} is from the entropy chain rule, where the transitions to \eqref{eq:CondEntrop210} and to~\eqref{eq:CondEntrop834} use the fact that conditioning reduces entropy, and where \eqref{eq:ReconEntroAll} is from the fact that the knowledge of $\{y^{n}, \tilde{y}^{n}, s^{n}_{0},\Omega^{n}\}$ allows for the reconstruction of  $\{z^{n}_{D,P},z^{n}_{N,P},z^{n}_{D,N},z^{n}_{N,D}, z^{n}_{D,D},z^{n}_{N,N}\}$ $($for example, knowing $\{y^{n}_{D,P}, \tilde{y}^{n}_{D,P}, s^{n}_{D,P},\Omega^{n}\}$, allows for reconstruction of $\{z^{n}_{D,P}\})$.

By adding \eqref{eq:2R1sum} and \eqref{eq:ReconEntroAll}, and dividing by $n$, we have
\begin{align}
& 2 R_1 + R_2 -  3 \epsilon_n \nonumber\\
&\leq  \frac{1}{n}\Bigl(h(y^{n}|\Omega^{n}) +h(\tilde{y}^{n}|\Omega^{n})  +  h(s^{n}_{0})     +  h(z^{n}_{P,P},z^{n}_{P,D},z^{n}_{P,N})  +  no(\log\rho) \Bigr)     \\
& \leq  2 \Bigl ( \sum_{\forall  (I_1,I_2) } \sum_{A_2\in \{  1, \alpha\} }  \sum_{A_1\in \{  1, \alpha\} }  A_1 \lambda_{ I_1,I_2}^{ A_1, A_2} \Bigr)  \log \rho
\nonumber \\  &\quad    +  \sum_{(I_1,I_2):I_1\neq P }  ( 1 -   \alpha ) \lambda_{ I_1,I_2}^{\alpha,  1} \log \rho
\nonumber \\  &\quad    +  \sum_{(I_1,I_2):I_1= P }  \sum_{A_2\in \{  1, \alpha\} }  \sum_{A_1\in \{  1, \alpha\} }  A_2 \lambda_{ I_1,I_2}^{ A_1, A_2} \log \rho
+  o(\log\rho)
\end{align}
and consequently have
\begin{align}
2d_1 + d_2 &\leq  2 \Bigl ( \sum_{(I_1,I_2)} \sum_{A_2\in \{  1, \alpha\} }  \sum_{A_1\in \{  1, \alpha\} }  A_1 \lambda_{ I_1,I_2}^{ A_1, A_2} \Bigr)
\nonumber \\  &\quad    +  \sum_{(I_1,I_2):I_1\neq P }  ( 1 -   \alpha ) \lambda_{ I_1,I_2}^{\alpha,  1}
\nonumber \\  &\quad    +  \sum_{(I_1,I_2):I_1=P }  \sum_{A_2\in \{  1, \alpha\} }  \sum_{A_1\in \{  1, \alpha\} }  A_2 \lambda_{ I_1,I_2}^{ A_1, A_2}  . \label{eq:2D1D2sum}
\end{align}

Similarly, exchanging the roles of user~1 and user~2, gives
\begin{align}
2d_2 + d_1 &\leq  2 \Bigl ( \sum_{\forall  I_1I_2 } \sum_{A_2\in \{  1, \alpha\} }  \sum_{A_1\in \{  1, \alpha\} }  A_2 \lambda_{ I_1,I_2}^{ A_1, A_2} \Bigr)
\nonumber \\  &\quad    +  \sum_{(I_1,I_2):I_2\neq P }  ( 1 -   \alpha ) \lambda_{ I_1,I_2}^{ 1,  \alpha}
\nonumber \\  &\quad    +  \sum_{(I_1,I_2):I_2= P }  \sum_{A_2\in \{  1, \alpha\} }  \sum_{A_1\in \{  1, \alpha\} }  A_1 \lambda_{ I_1,I_2}^{ A_1, A_2}  . \label{eq:2D2D1sum}
\end{align}

Consequently, summing up  the two bounds in \eqref{eq:2D1D2sum} and \eqref{eq:2D2D1sum} gives the following sum GDoF bound
\begin{align}
d_1 + d_2  &\leq \frac{1}{3}   \Bigl [  2 \Bigl ( \sum_{\forall  I_1I_2 } \sum_{A_2\in \{  1, \alpha\} }  \sum_{A_1\in \{  1, \alpha\} }  ( A_1+ A_2) \lambda_{ I_1,I_2}^{ A_1, A_2} \Bigr)
\nonumber \\  &\quad    +  \sum_{(I_1,I_2):I_1\neq P }  ( 1 -   \alpha ) \lambda_{ I_1,I_2}^{\alpha,  1}   +    \sum_{(I_1,I_2):I_2\neq P }  ( 1 -   \alpha ) \lambda_{ I_1,I_2}^{ 1,  \alpha}
\nonumber \\  &\quad    +  \!\!\!\sum_{(I_1,I_2):I_1= P }  \sum_{A_2\in \{  1, \alpha\} }  \sum_{A_1\in \{  1, \alpha\} }  A_2 \lambda_{ I_1,I_2}^{ A_1, A_2}  +\!\!\! \sum_{(I_1,I_2):I_2= P }  \sum_{A_2\in \{  1, \alpha\} }  \sum_{A_1\in \{  1, \alpha\} }  A_1 \lambda_{ I_1,I_2}^{ A_1, A_2}   \Bigr]
\end{align}
which, after some manipulation gives
\begin{align}
d_1 + d_2  &\leq   (1+\alpha)  (  \lambda_{P,P}^{\alpha, 1}  +  \lambda_{P,P}^{1, \alpha}   )
 +  \frac{3 + 2 \alpha}{3}  (   \lambda_{ P\leftrightarrow D}^{\alpha, 1}  +  \lambda_{ P\leftrightarrow D}^{1, \alpha}   )  +  \frac{3 + 2 \alpha}{3}  (  \lambda_{P\leftrightarrow N}^{\alpha, 1}  +  \lambda_{P\leftrightarrow N}^{1, \alpha}   )
\nonumber \\  &\quad +  \frac{3 +  \alpha}{3}  (  \lambda_{D,D}^{\alpha, 1}  +  \lambda_{D,D}^{1, \alpha}   )  +  \frac{3 +  \alpha}{3}  (  \lambda_{D\leftrightarrow N}^{\alpha, 1}  +  \lambda_{D\leftrightarrow N}^{1, \alpha}   ) +  \frac{3 +  \alpha}{3}  (  \lambda_{N,N}^{\alpha, 1}  +  \lambda_{N,N}^{1, \alpha}   )
\nonumber \\  &\quad   +  2  \lambda_{P,P}^{ 1, 1}  +  2\alpha  \lambda_{P,P}^{ \alpha, \alpha}
+ \frac{ 5 }{3}  \lambda_{P\leftrightarrow D}^{ 1, 1}  + \frac{ 5\alpha }{3} \lambda_{P\leftrightarrow D}^{ \alpha, \alpha}
+ \frac{ 5 }{3}  \lambda_{P\leftrightarrow N}^{ 1, 1}   + \frac{ 5\alpha }{3}  \lambda_{P\leftrightarrow N}^{ \alpha, \alpha}
\nonumber \\  &\quad   + \frac{ 4 }{3}  \lambda_{D,D}^{ 1, 1}  +  \frac{ 4\alpha }{3}  \lambda_{D,D}^{ \alpha, \alpha}
+ \frac{ 4 }{3}   \lambda_{D\leftrightarrow N}^{1, 1} + \frac{ 4\alpha }{3}  \lambda_{D\leftrightarrow N}^{\alpha, \alpha}
+ \frac{ 4 }{3}   \lambda_{N,N}^{ 1, 1} + \frac{ 4\alpha }{3}  \lambda_{N,N}^{ \alpha, \alpha} \label{eq:D1D2sum}.
\end{align}

\subsection{Proof for $d_1+ d_2 \leq d_{\sum}^{(4)}$  \ (cf. \eqref{eq:outerbound2})}

We continue with the proof of \eqref{eq:outerbound2}.
We first enhance the BC, by substituting delayed CSIT with perfect CSIT, i.e., by treating CSIT state $I_k = D$ as if it corresponded to $I_k = P$. We then transition to the compound BC by introducing a first imaginary user~$\tilde{1}$, and a second imaginary user~$\tilde{2}$.

User~$\tilde{1}$, which shares the same desired message $W_1$ as user~1, is supplied with a received signal that takes the form
\[    \tilde{y}^{n} = \bigl(  y^{n}_{P,P} ,  y^{n}_{P,D}, y^{n}_{D,P}, y^{n}_{D,D}, y^{n}_{P,N}, y^{n}_{D,N},\tilde{y}^{n}_{N,P}, \tilde{y}^{n}_{N,D}, \tilde{y}^{n}_{N,N} \bigr)\]
which means that user~1 and user~$\tilde{1}$ share the exact same received signal whenever $I_1\neq N$, while otherwise we only assume that user~$\tilde{1}$ has a received signal that is statistically identical to that of user~1, but not necessarily the same.

Similarly user~$\tilde{2}$, which shares the same desired message $W_2$ as user~2, is supplied with a received signal that takes the form
 \[    \tilde{z}^{n} = \bigl(  z^{n}_{P,P} ,  z^{n}_{D,P}, z^{n}_{P,D}, z^{n}_{D,D},  z^{n}_{N,P}, z^{n}_{N,D},  \tilde{z}^{n}_{P,N}, \tilde{z}^{n}_{D,N}, \tilde{z}^{n}_{N,N} \bigr)\]
which again means that user~2 and user~$\tilde{2}$ share the same received signal whenever $I_2\neq N$, while otherwise we only assume that user~$\tilde{2}$ has a received signal that is statistically identical to that of user~2, but not necessarily the same.

This latter stage does not further alter the capacity - compared to the previously \emph{enhanced} BC - since user~1 and user~$\tilde{1}$ have the same long-term decoding ability; similarly for user~2 and user~$\tilde{2}$.

Furthermore, whenever $(I_1,I_2)=(N,N)$ we can assume without an effect on the result, that the channel vectors $\gv_t, \tilde{\gv}_t, \tilde{\hv}_t, \hv_t$ are the same for all four users, i.e., $\gv_t =\tilde{\gv}_t = \tilde{\hv}_t= \hv_t$, ($\tilde{\gv}_t$ and $\tilde{\hv}_t$ for user~$\tilde{2}$ and user~$\tilde{1}$ respectively), since the capacity depends only on the marginals for the channels associated with $(I_1,I_2)=(N,N)$.

Additionally for any $t$ during which $(I_1,I_2)=(N,N)$, we define
\begin{align}  \label{eq:NNbarYsignalDef}
\bar{y}_t &=  \sqrt{\rho^{ \min \{A_{1,t}, A_{2,t} \}}} \hv^{\T}_t \xv_t  +   \bar{u}_t  \end{align}
where $\bar{u}_t $ is a unit-power AWGN random variable, where
\begin{align}
\sqrt{\rho^{A_{1,t}  -  \min \{A_{1,t}, A_{2,t} \} } } \bar{y}_t  &=  \underbrace{ \sqrt{\rho^{A_{1,t}}} \hv^{\T}_t \xv_t  +  u_t}_{ =  y_t   }       + \underbrace{\sqrt{\rho^{A_{1,t}  -  \min \{A_{1,t}, A_{2,t} \} } } \bar{u}_t   -u_t  }_{  \defeq \omega_t   }    \label{eq:NNrandom1}  \\
\sqrt{\rho^{A_{2,t}  -  \min \{A_{1,t}, A_{2,t} \} } } \bar{y}_t  &=  \underbrace{ \sqrt{\rho^{A_{2,t}}} \hv^{\T}_t \xv_t  +  v_t}_{ =  z_t   }       + \underbrace{\sqrt{\rho^{A_{2,t}  -  \min \{A_{1,t}, A_{2,t} \} } } \bar{u}_t   -v_t  }_{  \defeq \psi_t   }   \label{eq:NNrandom2}
\end{align}
and where the two new random variables $\omega_t, \psi_t$ have power
$$ \E|\omega_t|^2 \doteq \rho^{(A_{1,t}  -   A_{2,t})^{+} } $$ and $$\E|\psi_t|^2 \doteq \rho^{(A_{2,t}  -   A_{1,t})^{+} } .$$
The collection of all $\{ \bar{y}_t\}_t$ for all $t$ such that $(I_1,I_2)=(N,N)$, is denoted by $\bar{y}^{n}_{N,N}$, and similarly $\omega^{n}_{N,N}$ and $\psi^{n}_{N,N}$ respectively denote the set of $\{\omega_t\}_t$ and $\{\psi_t\}_t$ for all $t$ such that $(I_1,I_2)=(N,N)$.

Finally we provide each user with the observation $\bar{y}^{n}_{NN}$, to reach an enhanced compound BC.

At this point we have
\begin{align}
& nR_1-  n \epsilon_n \nonumber\\
& = H(W_1) - n \epsilon_n \nonumber\\
&=  H(W_1|\Omega^{n}) - n \epsilon_n \nonumber\\
&\leq I(W_1; y^{n}_{0},  y^{n}_{P,N}, y^{n}_{N,P}, y^{n}_{D,N}, y^{n}_{N,D}, y^{n}_{N,N},  \bar{y}^{n}_{N,N}|\Omega^{n})  \label{eq:gR1b0}\\
&= I(W_1; y^{n}_{0},  y^{n}_{P,N}, y^{n}_{N,P}, y^{n}_{D,N}, y^{n}_{N,D}, \bar{y}^{n}_{N,N}|\Omega^{n})   +   I(W_1; y^{n}_{N,N}| y^{n}_{0},  y^{n}_{P,N}, y^{n}_{N,P}, y^{n}_{D,N}, y^{n}_{N,D}, \bar{y}^{n}_{N,N},\Omega^{n})   \\
&= I(W_1; y^{n}_{0},  y^{n}_{P,N}, y^{n}_{N,P}, y^{n}_{D,N}, y^{n}_{N,D}, \bar{y}^{n}_{N,N}|\Omega^{n})   \nonumber\\&\quad +   \underbrace{ h( y^{n}_{N,N}| y^{n}_{0},  y^{n}_{P,N}, y^{n}_{N,P}, y^{n}_{D,N}, y^{n}_{N,D}, \bar{y}^{n}_{N,N},\Omega^{n}) }_{ \leq  h (y^{n}_{N,N}| \bar{y}^{n}_{N,N},\Omega^{n}) }-  \underbrace{ h( y^{n}_{N,N}| y^{n}_{0},  y^{n}_{P,N}, y^{n}_{N,P}, y^{n}_{D,N}, y^{n}_{N,D}, \bar{y}^{n}_{N,N},W_1,\Omega^{n}) }_{\geq  h( y^{n}_{N,N}| y^{n}_{0},  y^{n}_{P,N}, y^{n}_{N,P}, y^{n}_{D,N}, y^{n}_{N,D}, \bar{y}^{n}_{N,N},W_1,W_2,\Omega^{n})  \geq  no(\log \rho)  }   \\
&\leq  I(W_1; y^{n}_{0},  y^{n}_{P,N}, y^{n}_{N,P}, y^{n}_{D,N}, y^{n}_{N,D}, \bar{y}^{n}_{N,N}|\Omega^{n})   +   \underbrace{ h (y^{n}_{N,N}| \bar{y}^{n}_{N,N},\Omega^{n}) }_{ = h (\omega^{n}_{N,N}| \bar{y}^{n}_{N,N},\Omega^{n})  \leq  h (\omega^{n}_{N,N})  }+ no(\log \rho)    \label{eq:R1condEntropy}  \\
&\leq  I(W_1; y^{n}_{0},  y^{n}_{P,N}, y^{n}_{N,P}, y^{n}_{D,N}, y^{n}_{N,D}, \bar{y}^{n}_{N,N}|\Omega^{n})   +   h (\omega^{n}_{N,N}) + no(\log \rho)  \label{eq:recons2194} \\
&= h (\omega^{n}_{N,N}) + no(\log \rho)  + \underbrace{I(W_1; y^{n}_{0} |   y^{n}_{P,N}, y^{n}_{N,P}, y^{n}_{D,N}, y^{n}_{N,D}, \bar{y}^{n}_{N,N}, \Omega^{n})}_{ \leq  h( y^{n}_{0} ) +   no(\log \rho) }   \nonumber\\ & \quad +   I(W_1;   y^{n}_{P,N}, y^{n}_{N,P}, y^{n}_{D,N}, y^{n}_{N,D}, \bar{y}^{n}_{N,N} |\Omega^{n})     \\
&\leq h (\omega^{n}_{N,N}) +h( y^{n}_{0} )+ no(\log \rho)    +   I(W_1;   y^{n}_{P,N}, y^{n}_{N,P}, y^{n}_{D,N}, y^{n}_{N,D}, \bar{y}^{n}_{N,N} |\Omega^{n})    \label{eq:cond42814}  \\
&=  h (\omega^{n}_{N,N}) + h( y^{n}_{0} ) +  no(\log \rho)  +  I(W_1;   y^{n}_{P,N},y^{n}_{D,N},\bar{y}^{n}_{N,N} |\Omega^{n}) +   I(W_1;    y^{n}_{N,P}, y^{n}_{N,D} |y^{n}_{P,N},y^{n}_{D,N},\bar{y}^{n}_{N,N},\Omega^{n})   \\
&=  h (\omega^{n}_{N,N}) + \underbrace{h( y^{n}_{0} )}_{\leq n \Phi_{10} +  no(\log \rho) }+  no(\log \rho)  +  I(W_1;   y^{n}_{P,N},y^{n}_{D,N},\bar{y}^{n}_{N,N} |\Omega^{n})  \nonumber\\ &\quad  +    \underbrace{ I(W_1,W_2;    y^{n}_{N,P}, y^{n}_{N,D} |y^{n}_{P,N},y^{n}_{D,N},\bar{y}^{n}_{N,N},\Omega^{n})}_{\leq    h(  y^{n}_{N,P}, y^{n}_{N,D} ) +  no(\log \rho) \leq n \Phi_{11} +  no(\log \rho)} -  I(W_2;    y^{n}_{N,P}, y^{n}_{N,D} |W_1, y^{n}_{P,N},y^{n}_{D,N},\bar{y}^{n}_{N,N},\Omega^{n})   \\
&\leq  h (\omega^{n}_{N,N}) +  n \Phi_{10}  +n \Phi_{11} +  no(\log \rho)  \nonumber\\ &\quad +   I(W_1;   y^{n}_{P,N},y^{n}_{D,N},\bar{y}^{n}_{N,N} |\Omega^{n})     -  I(W_2;    y^{n}_{N,P}, y^{n}_{N,D} |W_1, y^{n}_{P,N},y^{n}_{D,N},\bar{y}^{n}_{N,N},\Omega^{n})   \label{eq:gR1boundf}
\end{align}
where
\[y^{n}_{0}  \defeq \bigl( y^{n}_{P,P} ,  y^{n}_{P,D}, y^{n}_{D,P}, y^{n}_{D,D} \bigr)  \]
where
\[     \Phi_{10}   \defeq  \bigl ( \sum_{ (I_1,I_2):I_1\neq N,  I_2\neq N  } \sum_{A_2\in \{  1, \alpha\} }  \sum_{A_1\in \{  1, \alpha\} }  A_1 \lambda_{ I_1,I_2}^{ A_1, A_2} \bigr)  \log \rho   \]
and
\[     \Phi_{11}   \defeq   \bigl( \sum_{(I_1,I_2) \in \{  (N,P), (N,D)\} } \sum_{A_2\in \{  1, \alpha\} }  \sum_{A_1\in \{  1, \alpha\} }  A_1 \lambda_{ I_1,I_2}^{ A_1, A_2} \bigr)  \log \rho   \]
where \eqref{eq:gR1b0} results from Fano's inequality, where the transition to \eqref{eq:R1condEntropy} uses the fact that conditioning reduces entropy and the fact that  $y^{n}_{N,N}$ can be reconstructed with errors up to noise level by using the knowledge of  $\{W_1,W_2,\Omega^{n}\}$, where \eqref{eq:recons2194} follows from the definition in \eqref{eq:NNrandom1} and from the fact that conditioning reduces entropy, and where \eqref{eq:cond42814} - \eqref{eq:gR1boundf} are derived using basic entropy rules.

Similarly  for user~$\tilde{1}$, we have
\begin{align}
& nR_1-  n \epsilon_n \nonumber\\
&\leq   h (\omega^{n}_{N,N}) +  n \Phi_{10}  +n \Phi_{11}  +  no(\log \rho)    \nonumber\\ &\quad  +   I(W_1;   y^{n}_{P,N},y^{n}_{D,N},\bar{y}^{n}_{N,N} |\Omega^{n})  -  I(W_2;    \tilde{y}^{n}_{N,P}, \tilde{y}^{n}_{N,D} |W_1, y^{n}_{P,N},y^{n}_{D,N},\bar{y}^{n}_{N,N},\Omega^{n}).  \label{eq:gR1boundfcomp}
\end{align}

Adding \eqref{eq:gR1boundf} and \eqref{eq:gR1boundfcomp}, gives
\begin{align}
& 2nR_1-  2 n \Phi_{10} - 2n \Phi_{11}  - no(\log \rho) - 2n \epsilon_n \nonumber\\
&\leq  2 h (\omega^{n}_{N,N}) + 2I(W_1;   y^{n}_{P,N},y^{n}_{D,N},\bar{y}^{n}_{N,N} |\Omega^{n}) - I(W_2;    y^{n}_{N,P}, y^{n}_{N,D} |W_1, y^{n}_{P,N},y^{n}_{D,N},\bar{y}^{n}_{N,N},\Omega^{n})  \nonumber\\ &\quad - I(W_2;    \tilde{y}^{n}_{N,P}, \tilde{y}^{n}_{N,D} |W_1, y^{n}_{P,N},y^{n}_{D,N},\bar{y}^{n}_{N,N},\Omega^{n}) \\
&=  2 h (\omega^{n}_{N,N}) + 2I(W_1;   y^{n}_{P,N},y^{n}_{D,N},\bar{y}^{n}_{N,N} |\Omega^{n})  \nonumber\\ &\quad \underbrace{ - h(  y^{n}_{N,P}, y^{n}_{N,D} |W_1, y^{n}_{P,N},y^{n}_{D,N},\bar{y}^{n}_{N,N},\Omega^{n}) - h( \tilde{y}^{n}_{N,P}, \tilde{y}^{n}_{N,D} |W_1, y^{n}_{P,N},y^{n}_{D,N},\bar{y}^{n}_{N,N},\Omega^{n})  }_{\leq -h(  y^{n}_{N,P}, y^{n}_{N,D},\tilde{y}^{n}_{N,P}, \tilde{y}^{n}_{N,D} |W_1, y^{n}_{P,N},y^{n}_{D,N},\bar{y}^{n}_{N,N},\Omega^{n})  }   \nonumber\\ &\quad  + \underbrace{h(y^{n}_{N,P}, y^{n}_{N,D} |W_2,W_1, y^{n}_{P,N},y^{n}_{D,N},\bar{y}^{n}_{N,N},\Omega^{n}) }_{=  no(\log \rho)  } +  \underbrace{h( \tilde{y}^{n}_{N,P}, \tilde{y}^{n}_{N,D} |W_2,W_1, y^{n}_{P,N},y^{n}_{D,N},\bar{y}^{n}_{N,N},\Omega^{n}) }_{=   no(\log \rho)  }  \nonumber\\
&\leq 2 h (\omega^{n}_{N,N}) + 2I(W_1;   y^{n}_{P,N},y^{n}_{D,N},\bar{y}^{n}_{N,N} |\Omega^{n}) - h(  y^{n}_{N,P}, y^{n}_{N,D},\tilde{y}^{n}_{N,P}, \tilde{y}^{n}_{N,D} |W_1, y^{n}_{P,N},y^{n}_{D,N},\bar{y}^{n}_{N,N},\Omega^{n})    \nonumber\\ &\quad +no(\log \rho)  \\
&= 2 h (\omega^{n}_{N,N}) + 2I(W_1;   y^{n}_{P,N},y^{n}_{D,N},\bar{y}^{n}_{N,N} |\Omega^{n}) - I(W_2;    y^{n}_{N,P}, y^{n}_{N,D},\tilde{y}^{n}_{N,P}, \tilde{y}^{n}_{N,D} |W_1, y^{n}_{P,N},y^{n}_{D,N},\bar{y}^{n}_{N,N},\Omega^{n})  \nonumber\\ &\quad  +\underbrace{ h( y^{n}_{N,P}, y^{n}_{N,D},\tilde{y}^{n}_{N,P}, \tilde{y}^{n}_{N,D} |W_2,W_1, y^{n}_{P,N},y^{n}_{D,N},\bar{y}^{n}_{N,N},\Omega^{n}) }_{=  no(\log \rho)  } +no(\log \rho)   \\
&=  2 h (\omega^{n}_{N,N}) + 2I(W_1;   y^{n}_{P,N},y^{n}_{D,N},\bar{y}^{n}_{N,N} |\Omega^{n}) - I(W_2;    y^{n}_{N,P}, \tilde{y}^{n}_{N,P}, y^{n}_{N,D},\tilde{y}^{n}_{N,D}, y^{n}_{P,N},y^{n}_{D,N},\bar{y}^{n}_{N,N} |W_1,\Omega^{n})  \nonumber\\ &\quad  + I(W_2;   y^{n}_{P,N},y^{n}_{D,N},\bar{y}^{n}_{N,N}|W_1,\Omega^{n}) + no(\log \rho)
\\
&=   2 h (\omega^{n}_{N,N}) + I(W_1;   y^{n}_{P,N},y^{n}_{D,N},\bar{y}^{n}_{N,N} |\Omega^{n}) - I(W_2;  y^{n}_{N,P},\tilde{y}^{n}_{N,P}, y^{n}_{N,D},\tilde{y}^{n}_{N,D}, y^{n}_{P,N},y^{n}_{D,N},\bar{y}^{n}_{N,N} |W_1,\Omega^{n})  \nonumber\\ &\quad  + \underbrace{I(W_1,W_2;   y^{n}_{P,N},y^{n}_{D,N},\bar{y}^{n}_{N,N}|\Omega^{n})}_{\leq  h( y^{n}_{P,N},y^{n}_{D,N},\bar{y}^{n}_{N,N})  +  no(\log \rho)  }  + no(\log \rho)\\
&\leq    h (\omega^{n}_{N,N}) + \underbrace{h (\omega^{n}_{N,N}) +   h( y^{n}_{P,N},y^{n}_{D,N},\bar{y}^{n}_{N,N})}_{ \leq  n \Phi_{12} +  no(\log \rho)  } +   I(W_1;   y^{n}_{P,N},y^{n}_{D,N},\bar{y}^{n}_{N,N} |\Omega^{n}) \nonumber\\ &\quad      - I(W_2;  y^{n}_{N,P},\tilde{y}^{n}_{N,P}, y^{n}_{N,D},\tilde{y}^{n}_{N,D}, y^{n}_{P,N},y^{n}_{D,N},\bar{y}^{n}_{N,N} |W_1,\Omega^{n})  +  no(\log \rho)\\
&\leq h (\omega^{n}_{N,N}) + n \Phi_{12} +  no(\log \rho)    +  I(W_1;   y^{n}_{P,N},y^{n}_{D,N},\bar{y}^{n}_{N,N} |\Omega^{n})  \nonumber\\ & \quad - I(W_2;     y^{n}_{N,P}, \tilde{y}^{n}_{N,P}, y^{n}_{N,D},\tilde{y}^{n}_{N,D}, y^{n}_{P,N},y^{n}_{D,N},\bar{y}^{n}_{N,N} |W_1,\Omega^{n})   \\
&= h (\omega^{n}_{N,N}) + n \Phi_{12} +  no(\log \rho)    +  I(W_1;   y^{n}_{P,N},y^{n}_{D,N},\bar{y}^{n}_{N,N} |\Omega^{n})  \nonumber\\ & \quad  - I(W_2;    y^{n}_{N,P}, \tilde{y}^{n}_{N,P}, s^{n}_{N,P},  y^{n}_{N,D}, \tilde{y}^{n}_{N,D}, s^{n}_{N,D},  y^{n}_{P,N},y^{n}_{D,N},\bar{y}^{n}_{N,N}   |W_1,\Omega^{n})  \nonumber\\ & \quad +  \underbrace{I(W_2;    s^{n}_{N,P}, s^{n}_{N,D}  | y^{n}_{N,P}, \tilde{y}^{n}_{N,P}, y^{n}_{N,D},\tilde{y}^{n}_{N,D}, y^{n}_{P,N},y^{n}_{D,N},\bar{y}^{n}_{N,N},  W_1,\Omega^{n})}_{\leq  h( s^{n}_{N,P}, s^{n}_{N,D} ) + no(\log \rho) }    \label{eq:Ssignal}   \\
&\leq  h (\omega^{n}_{N,N}) + n \Phi_{12}    +  I(W_1;   y^{n}_{P,N},y^{n}_{D,N},\bar{y}^{n}_{N,N} |\Omega^{n})  \nonumber\\ & \quad  - I(W_2;    y^{n}_{N,P}, \tilde{y}^{n}_{N,P}, s^{n}_{N,P},  y^{n}_{N,D}, \tilde{y}^{n}_{N,D}, s^{n}_{N,D},  y^{n}_{P,N},y^{n}_{D,N},\bar{y}^{n}_{N,N}   |W_1,\Omega^{n})  +  h( s^{n}_{N,P}, s^{n}_{N,D} ) + no(\log \rho)   \\
& =  h (\omega^{n}_{N,N}) +  n \Phi_{12}   +  I(W_1;   y^{n}_{P,N},y^{n}_{D,N},\bar{y}^{n}_{N,N} |\Omega^{n}) + h( s^{n}_{N,P}, s^{n}_{N,D} ) +no(\log \rho)  \nonumber\\ & \quad  - I(W_2;    y^{n}_{N,P}, \tilde{y}^{n}_{N,P}, s^{n}_{N,P},  z^{n}_{N,P},  y^{n}_{N,D}, \tilde{y}^{n}_{N,D}, s^{n}_{N,D}, z^{n}_{N,D}, y^{n}_{P,N},y^{n}_{D,N},\bar{y}^{n}_{N,N}   |W_1,\Omega^{n})        \label{eq:EntropyEq}  \\
& \leq h (\omega^{n}_{N,N}) +   n \Phi_{12}   +  I(W_1;   y^{n}_{P,N},y^{n}_{D,N},\bar{y}^{n}_{N,N} |\Omega^{n}) + h( s^{n}_{N,P}, s^{n}_{N,D} ) + no(\log \rho)\nonumber\\ & \quad  - I(W_2;   z^{n}_{N,P}, z^{n}_{N,D},\bar{y}^{n}_{N,N}   |W_1,\Omega^{n})        \label{eq:ConEntropy}  \\
& \leq  h (\omega^{n}_{N,N}) +  n \Phi_{12}   +  I(W_1;   W_2, y^{n}_{P,N},y^{n}_{D,N},\bar{y}^{n}_{N,N} |\Omega^{n}) + h( s^{n}_{N,P}, s^{n}_{N,D} ) + no(\log \rho)  \nonumber\\ & \quad  - I(W_2;   z^{n}_{N,P}, z^{n}_{N,D},\bar{y}^{n}_{N,N}   |W_1,\Omega^{n})         \\
& =  h (\omega^{n}_{N,N}) +  n \Phi_{12}   +  I(W_1;  y^{n}_{P,N},y^{n}_{D,N},\bar{y}^{n}_{N,N} | W_2, \Omega^{n}) + \underbrace{h( s^{n}_{N,P}, s^{n}_{N,D} )}_{ \leq n \Phi_{13} +  no(\log \rho)   } +no(\log \rho) \nonumber\\ & \quad  - I(W_2;   z^{n}_{N,P}, z^{n}_{N,D},\bar{y}^{n}_{N,N}   |W_1,\Omega^{n})    \\
& \leq   \underbrace{h (\omega^{n}_{N,N})}_{\leq n \Phi_{14} +  no(\log \rho)   } + n \Phi_{12}   +  I(W_1;  y^{n}_{P,N},y^{n}_{D,N},\bar{y}^{n}_{N,N} | W_2, \Omega^{n}) + n \Phi_{13} +no(\log \rho)    \nonumber\\ & \quad - I(W_2;   z^{n}_{N,P}, z^{n}_{N,D},\bar{y}^{n}_{N,N}   |W_1,\Omega^{n})    \\
& \leq   n \Phi_{14} + n \Phi_{12}   +  I(W_1;  y^{n}_{P,N},y^{n}_{D,N},\bar{y}^{n}_{N,N} | W_2, \Omega^{n}) + n \Phi_{13} +no(\log \rho)    \nonumber\\ & \quad - I(W_2;   z^{n}_{N,P}, z^{n}_{N,D},\bar{y}^{n}_{N,N}   |W_1,\Omega^{n})    \label{eq:2R1sumboundf}
\end{align}
where
\[     \Phi_{12}   \defeq  \bigl ( \sum_{(I_1,I_2) : I_2 = N } \sum_{A_2\in \{  1, \alpha\} }  \sum_{A_1\in \{  1, \alpha\} }  A_1 \lambda_{ I_1,I_2}^{ A_1, A_2} \bigr)  \log \rho   \]
\[     \Phi_{13}   \defeq  \bigl ( \sum_{(I_1,I_2) \in \{ ( N,P), (N,D)\} } (1 - \alpha) \lambda_{ I_1,I_2}^{\alpha, 1} \bigr)  \log \rho   \]
\[     \Phi_{14}   \defeq    (1 - \alpha) \lambda_{ N,N}^{1, \alpha}   \log \rho   \]
where $s^{n}_{N,P}$ and $z^{n}_{N,D}$ (cf. \eqref{eq:Ssignal}) are defined in \eqref{eq:SsignalDef}. Furthermore~\eqref{eq:EntropyEq} is from the fact that the knowledge of $\{y^{n}_{N,P}, \tilde{y}^{n}_{N,P}, s^{n}_{N,P},y^{n}_{N,D}, \tilde{y}^{n}_{N,D}, s^{n}_{N,D}, \Omega^{n}\}$ implies the knowledge of $z^{n}_{N,P}$ and $z^{n}_{N,D}$ (cf. \eqref{eq:SsignalDef}). Most of the above steps are based on basic entropy rules.

Similarly, considering user~2 and user~$\tilde{2}$, we have
\begin{align}
& 2nR_2-  2 n\Phi_{20} - 2n\Phi_{21}  - no(\log \rho) - 2n \epsilon_n \nonumber\\
& \leq   n \Phi_{24} +n \Phi_{22}   +  I(W_2;  z^{n}_{N,P},z^{n}_{N,D},\bar{y}^{n}_{N,N} | W_1, \Omega^{n}) +n \Phi_{23}+ no(\log \rho)  \nonumber\\& \quad - I(W_1;   y^{n}_{P,N}, y^{n}_{D,N},\bar{y}^{n}_{N,N}   |W_2,\Omega^{n})    \label{eq:2R2sumboundf}
\end{align}
where
\[     \Phi_{20}   \defeq  \bigl ( \sum_{(I_1,I_2):I_1\neq N,I_2\neq N } \sum_{A_2\in \{  1, \alpha\} }  \sum_{A_1\in \{  1, \alpha\} }  A_2 \lambda_{ I_1,I_2}^{ A_1, A_2} \bigr)  \log \rho   \]
\[     \Phi_{21}   \defeq   \bigl( \sum_{(I_1,I_2) \in \{  (P,N), (D,N)\} } \sum_{A_2\in \{  1, \alpha\} }  \sum_{A_1\in \{  1, \alpha\} }  A_2 \lambda_{ I_1,I_2}^{ A_1, A_2} \bigr)  \log \rho   \]
\[     \Phi_{22}   \defeq  \bigl ( \sum_{(I_1,I_2):I_1=N } \sum_{A_2\in \{  1, \alpha\} }  \sum_{A_1\in \{  1, \alpha\} }  A_2 \lambda_{ I_1,I_2}^{ A_1, A_2} \bigr)  \log \rho   \]
\[     \Phi_{23}   \defeq  \bigl ( \sum_{(I_1,I_2) \in \{ ( P,N), (D,N)\} } (1 - \alpha) \lambda_{ I_1,I_2}^{ 1, \alpha} \bigr)  \log \rho   \]
\[     \Phi_{24}   \defeq    (1 - \alpha) \lambda_{ N,N}^{\alpha, 1}   \log \rho.   \]

Finally,  combining \eqref{eq:2R1sumboundf} and \eqref{eq:2R2sumboundf}, gives
\begin{align}
& d_1  +   d_2    \nonumber \\
& \leq  \frac{1}{2\log \rho } \Bigl[  2 \Phi_{10} + 2\Phi_{11}   +   \Phi_{12}  + \Phi_{13}    +\Phi_{14}   +    2 \Phi_{20} + 2\Phi_{21}   +   \Phi_{22}  + \Phi_{23}  +\Phi_{24}       \Bigr]    \nonumber\\
& =  \frac{1}{2 } \Bigl[   2 \bigl ( \sum_{(I_1,I_2):I_1\neq N,I_2\neq N } \sum_{A_2\in \{  1, \alpha\} }  \sum_{A_1\in \{  1, \alpha\} }  A_1 \lambda_{ I_1,I_2}^{ A_1, A_2} \bigr)   \nonumber\\ &\quad  + 2 \bigl( \sum_{(I_1,I_2) \in \{  (N,P), (N,D)\} } \sum_{A_2\in \{  1, \alpha\} }  \sum_{A_1\in \{  1, \alpha\} }  A_1 \lambda_{ I_1,I_2}^{ A_1, A_2} \bigr)   \nonumber\\ &\quad +   \sum_{(I_1,I_2):I_2 = N } \sum_{A_2\in \{  1, \alpha\} }  \sum_{A_1\in \{  1, \alpha\} }  A_1 \lambda_{ I_1,I_2}^{ A_1, A_2} \nonumber\\ &\quad  +  \sum_{(I_1,I_2) \in \{ ( N,P), (N,D)\} } (1 - \alpha) \lambda_{ I_1,I_2}^{\alpha, 1}     +  (1 - \alpha) \lambda_{ N,N}^{1, \alpha}
\nonumber\\ &\quad +    2 \bigl ( \sum_{(I_1,I_2):I_1\neq N,I_2\neq N  } \sum_{A_2\in \{  1, \alpha\} }  \sum_{A_1\in \{  1, \alpha\} }  A_2 \lambda_{ I_1,I_2}^{ A_1, A_2} \bigr)     \nonumber\\ &\quad    + 2 \bigl( \sum_{(I_1,I_2) \in \{  (P,N), (D,N)\} } \sum_{A_2\in \{  1, \alpha\} }  \sum_{A_1\in \{  1, \alpha\} }  A_2 \lambda_{ I_1,I_2}^{ A_1, A_2} \bigr)    \nonumber\\ &\quad   +   \sum_{(I_1,I_2):I_1 = N } \sum_{A_2\in \{  1, \alpha\} }  \sum_{A_1\in \{  1, \alpha\} }  A_2 \lambda_{ I_1,I_2}^{ A_1, A_2}   \nonumber\\ &\quad   +   \sum_{(I_1,I_2) \in \{  (P,N), (D,N)\} } (1 - \alpha) \lambda_{ I_1,I_2}^{ 1, \alpha}   + (1 - \alpha) \lambda_{ N,N}^{\alpha, 1}    \Bigr] \nonumber\\
& =     \sum_{(I_1,I_2):I_1\neq N,I_2\neq N  }  \bigl( 1+ \alpha \bigr)\bigl( \lambda_{ I_1,I_2}^{ 1, \alpha} +\lambda_{ I_1,I_2}^{\alpha, 1} \bigr)  \nonumber\\ &\quad   + \sum_{(I_1,I_2) \in \{ ( N,P), (P,N), (N,D), (D,N)\} }   \frac{2+ \alpha }{2} \bigl(\lambda_{ I_1,I_2}^{ 1, \alpha} +\lambda_{ I_1,I_2}^{\alpha, 1} \bigr)   +   \bigl( \lambda_{ N,N}^{1, \alpha} +\lambda_{ N,N}^{\alpha, 1} \bigr)
 \nonumber\\ &\quad  + \sum_{(I_1,I_2):I_1\neq N,I_2\neq N  }  \bigl( 2 \lambda_{ I_1,I_2}^{1, 1} + 2\alpha \lambda_{ I_1,I_2}^{\alpha, \alpha}  \bigr)
 \nonumber\\ &\quad + \sum_{(I_1,I_2) \in \{  (N,P), (P,N), (N,D), (D,N)\} }   \bigl( \frac{3}{2} \lambda_{ I_1,I_2}^{1, 1} + \frac{3\alpha}{2} \lambda_{ I_1,I_2}^{\alpha, \alpha}  \bigr)     +  (   \lambda_{ N,N}^{ 1, 1} +\alpha\lambda_{ N,N}^{\alpha, \alpha} )   \nonumber\\
 & =      \bigl(1+ \alpha \bigr)\bigl(\lambda_{P,P}^{1, \alpha} +\lambda_{P,P}^{\alpha, 1} \bigr)  +   \bigl(1+ \alpha \bigr)\bigl(\lambda_{ P\leftrightarrow D}^{1, \alpha} +\lambda_{ P\leftrightarrow D}^{\alpha, 1} \bigr) +   \bigl(1+ \alpha \bigr)\bigl(\lambda_{D,D}^{1, \alpha} +\lambda_{D,D}^{\alpha, 1} \bigr)
 \nonumber\\ &\quad   + \frac{2+ \alpha }{2}\bigl(\lambda_{P\leftrightarrow N}^{1, \alpha} +\lambda_{P\leftrightarrow N}^{\alpha, 1} \bigr) + \frac{2+ \alpha }{2}\bigl(\lambda_{D\leftrightarrow N}^{1, \alpha} +\lambda_{D\leftrightarrow N}^{\alpha, 1} \bigr)   +  \bigl( \lambda_{ N,N}^{ 1, \alpha} +\lambda_{ N,N}^{ \alpha, 1} \bigr)
 \nonumber\\ &\quad  +  \bigl(2 \lambda_{P,P}^{ 1, 1} + 2\alpha \lambda_{P,P}^{ \alpha, \alpha}\bigr) +  \bigl(2 \lambda_{P\leftrightarrow D}^{ 1, 1} + 2\alpha \lambda_{P\leftrightarrow D}^{ \alpha, \alpha} \bigr) +  \bigl(2 \lambda_{D,D}^{ 1, 1} + 2\alpha \lambda_{D,D}^{ \alpha, \alpha} \bigr)
\nonumber\\ &\quad+  \bigl( \frac{3}{2} \lambda_{P\leftrightarrow N}^{ 1, 1} + \frac{3\alpha}{2} \lambda_{P\leftrightarrow N}^{ \alpha, \alpha}  \bigr)   +  \bigl( \frac{3}{2} \lambda_{D\leftrightarrow N}^{1, 1} + \frac{3\alpha}{2} \lambda_{D\leftrightarrow N}^{\alpha, \alpha}  \bigr)     +    \bigl(\lambda_{N,N}^{ 1, 1} +\alpha\lambda_{N,N}^{ \alpha, \alpha} \bigr)
\end{align}
which completes the proof.

\section{Appendix - schemes}

\begin{table}
\caption{Summary of schemes}
\begin{center}
{\renewcommand{\arraystretch}{1.7}
\begin{tabular}{|c|c|c|c|c|}
  \hline
     Scheme $\#$        &  Section $\#$  & CSIT, topology  &  achieved $d_{\sum} $   &  for Proposition $\#$  \\
   \hline
 1   & \ref{sec:ND1a-PN1a}  &   $\lambda_{1,\alpha} = 1$     &    $1+ \frac{\alpha}{2} $    &  Proposition~\ref{prop:simpleFB-fixedTopology}  \\
  &   &   $\lambda_{N,D} = \lambda_{P,N} = 1/2$         &  optimal     &  \\
    \hline
2   & \ref{sec:PD1aNN1a}  &   $\lambda_{1,\alpha} = 1
$   &    $1+ \frac{\alpha}{2} $    &  Proposition~\ref{prop:simpleFB-fixedTopology}  \\
  &   &   $\lambda_{P,D} = \lambda_{N,N} = 1/2$         &  optimal     &  \\
    \hline
 3  &   \ref{sec:DD1a}    &   $\lambda_{1,\alpha} = 1$       &    $1+  \frac{ \alpha^2}{2+ \alpha }  $    &  Proposition~\ref{prop:DD1a} \\
  &   &   $\lambda_{D, D} = 1$         &       &  \\
    \hline
 4   & \ref{sec:DD1a-DDa1}   &   $\lambda_{1,\alpha} = \lambda_{\alpha,1} = 1/2$    &    $1+ \frac{\alpha}{3} $    &  Proposition~\ref{prop:DD1a-DDa1} \\
  &   &   $\lambda_{D, D} = 1$         &   optimal    &  \\
    \hline
5  &   \ref{sec:PNNP-any}   &    any $\lambda_{1,\alpha} + \lambda_{\alpha,1} = 1$      &    $1+ \frac{\alpha}{2} $    &  Propositions~\ref{prop:simpleFB-fixedTopology}, \ref{prop:PNgeneral-altTopology}, \ref{prop:simpleFB-2Topologies} \\
  &   &   $\lambda_{P,N}=\lambda_{N,P} = 1/2$         & optimal      &  \\
    \hline
  MAT & \ref{sec:MATorig-DD1a}   &   $\lambda_{D, D}^{ 1,\alpha } =1$   &  $\frac{2(1 + \alpha)}{3} $  &  -  \\
  &   &            & sub-optimal      &  \\
    \hline
    \end{tabular}
}
\end{center}
\label{tab:schsummary}
\end{table}

We proceed to design the topological signal management schemes for the different topology and feedback scenarios (see Table~\ref{tab:schsummary} for a summary).
In what follows, we will generally associate the use of symbol $a$ to denote a private symbol for user 1, while we will associate symbol $b$ to denote a private symbol for user~2, and symbol $c$ to denote a common symbol meant for both users. We will also use $P^{(q)} \defeq \E |q|^2$ to denote the average power of some symbol $q$, and will use $r^{(q)}$ to denote the pre-log factor of the number of bits $[r^{(q)}\log \rho- o(\log \rho)]$ carried by symbol $q$.
In the interest of brevity, we will on occasion neglect the additive noise terms, without an effect on the GDoF analysis.

\subsection{ TSM scheme for $\lambda_{N,D}^{1,\alpha}=\lambda_{P,N}^{1,\alpha}= 1/2$ achieving the optimal sum GDoF $1 + \alpha/2$   \label{sec:ND1a-PN1a}}

For the setting of $\lambda_{N,D}^{1,\alpha}=\lambda_{P,N}^{1,\alpha}= 1/2$, the proposed scheme consists of two channel uses, which, without loss of generality, are assumed here to be consecutive. During the first channel use, $t = 1$, the feedback-and-topology state is $(I_1,I_2,A_1,A_2)= (N, D,  1 ,  \alpha)$, while during the second channel use, $t = 2$, the feedback-and-topology state is $(I_1,I_2,A_1,A_2)= (P, N,  1 ,  \alpha)$.

\begin{figure}
\centering
\includegraphics[width=12cm]{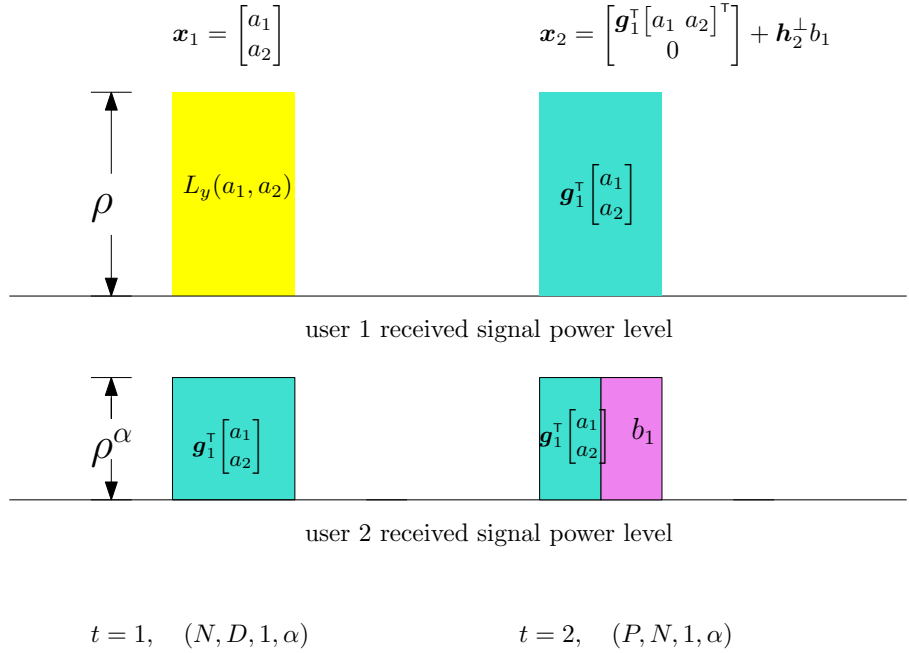}
\caption{ Illustration of received signal power level for the TSM scheme for $\lambda_{N,D}^{1,\alpha}=\lambda_{P,N}^{1,\alpha}= 1/2$ . }
\label{fig:TSMPDNNX5}
\end{figure}

At time $t=1$ there is no CSIT, and the transmitter sends (see Figure~\ref{fig:TSMPDNNX5})
\begin{align}
\xv_{1} =     \Bmatrix{a_1  \\  a_2 }
\end{align}
where $a_1$, $a_2$ are symbols meant for user 1, with
\begin{equation}
\begin{array}{cc}
P^{(a_1)} \doteq  1,  &  r^{(a_1)} = 1  \\
P^{(a_2)} \doteq  1,  &  r^{(a_2)} = 1
\end{array}
\end{equation}
resulting in received signals of the form
\begin{align}
  y_1&=   \underbrace{\sqrt{\rho}    \hv^\T_1 \Bmatrix{a_1  \\  a_2 } }_{\rho} +\underbrace{u_1}_{\rho^0} \label{eq:sch17777y1}\\
  z_1	&= \underbrace{ \sqrt{\rho^{\alpha}}\gv^\T_1 \Bmatrix{a_1  \\  a_2 } }_{\rho^{\alpha}} +\underbrace{v_1}_{\rho^0}   \label{eq:sch199777y2}
\end{align}
where under each term we noted the order of the summand's average power.
One can briefly note that the unintended interference is naturally attenuated due to the weak link.

At time $t=2$, the transmitter has knowledge of $\gv_1$ (delayed feedback) and of $\hv_2$ (current feedback). As a result, the transmitter reconstructs $\gv^\T_1 \Bmatrix{a_1  \  a_2 }^\T$ and sends
\begin{align}
\xv_{2} =    \Bmatrix{ \gv^\T_1 \Bmatrix{a_1  \  a_2 }^\T     \\  0  }    +   \hv_2^\bot  b_1
\end{align}
where $b_1$ is meant for user~2, and where
\begin{equation}
\begin{array}{cc}
P^{(b_1)} \doteq  1,  &  r^{(b_1)} = \alpha.
\end{array}
\end{equation}
Then the processed (normalized) received signals take the form
\begin{align}
  y_2 /h_{2,1}    &=   \underbrace{\sqrt{\rho}   \gv^\T_1 \Bmatrix{a_1  \\  a_2 }}_{\rho}  +\underbrace{\frac{u_2}{h_{2,1}}}_{\rho^0}   \label{eq:sch1y1384323}\\
  z_2	/ g_{2,1} &= \underbrace{ \sqrt{\rho^{\alpha}}\gv^\T_1 \Bmatrix{a_1  \\  a_2 }}_{\rho^{\alpha}}     +      \underbrace{ \sqrt{\rho^{\alpha}}  \frac{\gv^\T_2\hv_2^\bot}{ g_{2,1} }  b_1 }_{\rho^{\alpha}} +\underbrace{\frac{v_2}{g_{2,1}}}_{\rho^0}    \label{eq:sch1y2322123}
\end{align}
where $h_{t,1} \defeq \hv^\T_t \Bmatrix{1 \  0 }^\T $,   $g_{t,1} \defeq \gv^\T_t \Bmatrix{1 \  0 }^\T$, and where the normalized noise power (of $\frac{u_2}{h_{2,1}}$ and $\frac{v_2}{g_{2,1}}$) is noted to be \emph{typically} bounded, since $Pr(|h_{2,1}|^2 \leq \rho^{-\epsilon}) \doteq Pr(|g_{2,1}|^2 \leq \rho^{-\epsilon}) \doteq \rho^{-\epsilon}$ for arbitrarily small positive $\epsilon$.

At this point, it is easy to see that user~1 can recover $a_1, a_2$ at the declared rates, by MIMO decoding based on \eqref{eq:sch17777y1},  \eqref{eq:sch1y1384323}, while user~2 can recover $b_1$ by employing interference cancelation based on \eqref{eq:sch199777y2},  \eqref{eq:sch1y2322123}.  This provides for the optimal sum GDoF $d_{\sum}=1 + \alpha/2$.

\subsection{ TSM scheme for $\lambda_{P,D}^{1,\alpha}= \lambda_{N,N}^{1,\alpha}= 1/2$, achieving the optimal sum GDoF $1 + \alpha/2$   \label{sec:PD1aNN1a}}

For the setting where $\lambda_{P,D}^{1,\alpha}
= \lambda_{N,N}^{1,\alpha}= 1/2$, the proposed scheme has two channel uses. Again without loss of generality, we assume that during $t = 1$ the state is $(I_1,I_2,A_1,A_2)= (P, D,  1 ,  \alpha)$, while during $t = 2$, the state is $(I_1,I_2,A_1,A_2)= (N, N,  1 ,  \alpha)$.

\begin{figure}
\centering
\includegraphics[width=12cm]{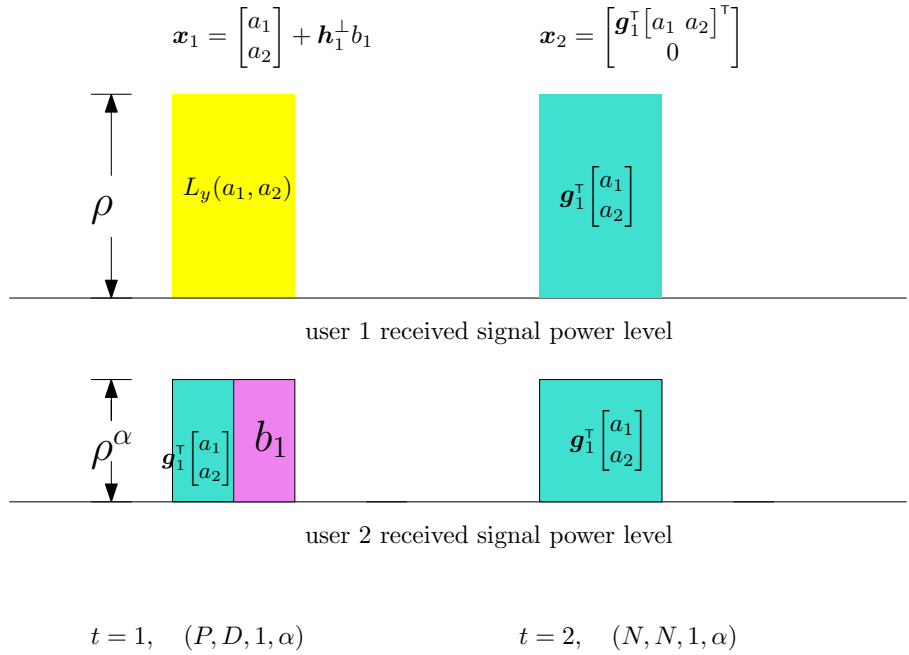}
\caption{ Illustration of received signal power level in TSM scheme for $\lambda_{P,D}^{1,\alpha}= \lambda_{N,N}^{1,\alpha}= 1/2$. }
\label{fig:TSMPDNNX4}
\end{figure}

At $t=1$ the transmitter knows $\hv_1$ (current CSIT) and sends (see Figure~\ref{fig:TSMPDNNX4})
\begin{align}
\xv_{1} =     \Bmatrix{a_1  \\  a_2 }  +  \hv_1^\bot  b_1
\end{align}
where $a_1$, $a_2$ are unit-power symbols meant for user 1, $b_1$ is a unit-power symbol  meant for user~2, and where
\begin{equation}\label{eq:RPX1837101}
\begin{array}{ccc}
 r^{(a_1)} = 1,   &  r^{(a_2)} = 1,  &  r^{(b_1)} = \alpha.
\end{array}
\end{equation}
Then the received signals (in their noiseless form) are
\begin{align}
  y_1&=   \underbrace{\sqrt{\rho}    \hv^\T_1 \Bmatrix{a_1  \\  a_2 } }_{\rho} \label{eq:sch1987y1}\\
  z_1	&= \underbrace{ \sqrt{\rho^{\alpha}}\gv^\T_1 \Bmatrix{a_1  \\  a_2 } }_{\rho^{\alpha}}  +\underbrace{ \sqrt{\rho^{\alpha} }\gv^\T_1 \hv_1^\bot b_1}_{\rho^{\alpha}}    \label{eq:sch1997y2}
\end{align}
where in the above we omitted the noise, for brevity and without an effect to the derived DoF expressions.

At $t=2$ $((I_1,I_2,A_1,A_2)= (N, N,  1 ,  \alpha))$, the transmitter knows $\gv_1$ (delayed CSIT), reconstructs $\gv^\T_1 \Bmatrix{a_1  \\  a_2 }$, and sends
\begin{align}
\xv_{2} =    \Bmatrix{ \gv^\T_1 \Bmatrix{a_1  \  a_2 }^\T   \\  \\  0  }.
\end{align}
After normalization, the received signals (in their noiseless form) are
\begin{align}
  y_2 / h_{2,1}    &=   \underbrace{\sqrt{\rho}   \gv^\T_1 \Bmatrix{a_1  \\  a_2 }}_{\rho}     \label{eq:sch1y1fad323}\\
  z_2	/ g_{2,1}&= \underbrace{ \sqrt{\rho^{\alpha}}\gv^\T_1 \Bmatrix{a_1  \\  a_2 }}_{\rho^{\alpha}}.      \label{eq:sch1y23afda123}
\end{align}

One can now easily see that, user~1 can MIMO decode $a_1, a_2$ based on \eqref{eq:sch1987y1} and \eqref{eq:sch1y1fad323}, while user~2 can recover $b_1$
by employing interference cancelation based on \eqref{eq:sch1997y2} and \eqref{eq:sch1y23afda123} (see also Figure~\ref{fig:TSMPDNNX4}).  This achieves the optimal sum GDoF $d_{\sum}=1 + \alpha/2$.

\vspace{3pt}

\subsection{TSM scheme for the case with
$\lambda_{D, D}^{ 1,\alpha }=1$  \label{sec:DD1a}}

\begin{figure}
\centering
\includegraphics[width=10cm]{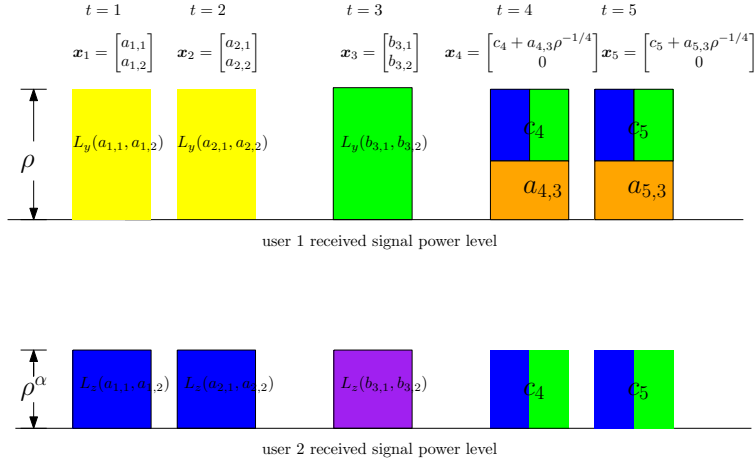}
\caption{ Received signal power level illustration for the proposed TSM scheme: The case with $\lambda_{D,D}^{1,\alpha}=1$ and $\alpha=1/2$. }
\label{fig:TSMsignalDD1a}
\end{figure}

The proposed scheme has three phases, of respective durations $T_1,T_2,T_3$ channel uses\footnote{Here we assume that $\alpha$ is a rational number, an assumption which automatically allows $T_1,T_2,T_3$ to be integer valued. The case of irrational $\alpha$ can be handled with minor modifications to the scheme.},\footnote{As a clarifying example, when $\alpha=1/2$, the phase durations are $T_1=2, \ T_2=1, \  T_3=2$ (see Figure~\ref{fig:TSMsignalDD1a}).}
\begin{align} \label{eq:phaseduration}
T_2/\alpha = T_1 = T_3
\end{align}
which - as we will see later on - are chosen so that the amounts of side information, at user~1 and user~2, are properly balanced.

\subsubsection{Phase~1}  When $t=1,2,\cdots,T_1$,  the transmitter sends
\begin{align}
\xv_{t} =   \Bmatrix{ a_{t,1} \\  a_{t,2} }
\end{align}
where $a_{t,1}$ and $a_{t,2}$ are unit-power symbols meant for user 1, and where
\begin{equation}  \label{eq:TSMDD1aPowerrate1}
\begin{array}{cc}
   r^{(a_{t,1})} = 1,  &   r^{(a_{t,2})} = \alpha.
\end{array}
\end{equation}
The received signals then take the form
\begin{align}
  y_t&=   \underbrace{\sqrt{\rho}    \hv^\T_t \Bmatrix{ a_{t,1} \\  a_{t,2} }   }_{\rho}   +  \underbrace{ u_t }_{\rho^0}  \label{eq:newMATy1}\\
  z_t	&= \underbrace{ \sqrt{\rho^{\alpha}}\gv^\T_t \Bmatrix{ a_{t,1} \\  a_{t,2} }   }_{\rho^{\alpha}}   +   \underbrace{ v_t }_{\rho^0}    = \sqrt{\rho^{\alpha}}  L_{z} (a_{t,1},  a_{t,2} )  +   v_t \label{eq:newMATy2}
\end{align}
where $L_{z} (a_{t,1},  a_{t,2} )\defeq \gv^\T_t \Bmatrix{ a_{t,1} \\  a_{t,2} }$ represents interference at the second receiver.

\subsubsection{Phase~2}  When $t=T_1+ 1, \cdots,T_1 + T_2$,  the transmitter sends
\begin{align}
\xv_{t} =     \Bmatrix{ b_{t,1} \\  b_{t,2} }
\end{align}
where  $b_{t,1},b_{t,2}$ are unit-power symbols meant for user 2, and where
\begin{equation} \label{eq:TSMDD1aPowerrate2}
\begin{array}{cc}
  r^{(b_{t,1})} = 1, &  r^{(b_{t,2})} = \alpha
\end{array}
\end{equation}
resulting in received signals of the form
\begin{align}
  y_t&=   \underbrace{ \sqrt{\rho}\hv^\T_t \Bmatrix{ b_{t,1} \\  b_{t,2} }   }_{\rho}     +  \underbrace{ u_t }_{\rho^0} =  \sqrt{\rho}   L_{y} (b_{t,1},  b_{t,2})  +u_t \label{eq:newMATy102}\\
  z_t	&=  \underbrace{\sqrt{\rho^{\alpha}}    \gv^\T_t \Bmatrix{ b_{t,1}   \\  b_{t,2} }   }_{\rho^{\alpha}}     +   \underbrace{ v_t }_{\rho^0}      \label{eq:newMATy202}
\end{align}
where $L_{y} (b_{t,1},  b_{t,2}) \defeq \hv^\T_t \Bmatrix{ b_{t,1} \\  b_{t,2} }$ represents interference at the first receiver.

 \begin{figure}
\centering
\includegraphics[width=14cm]{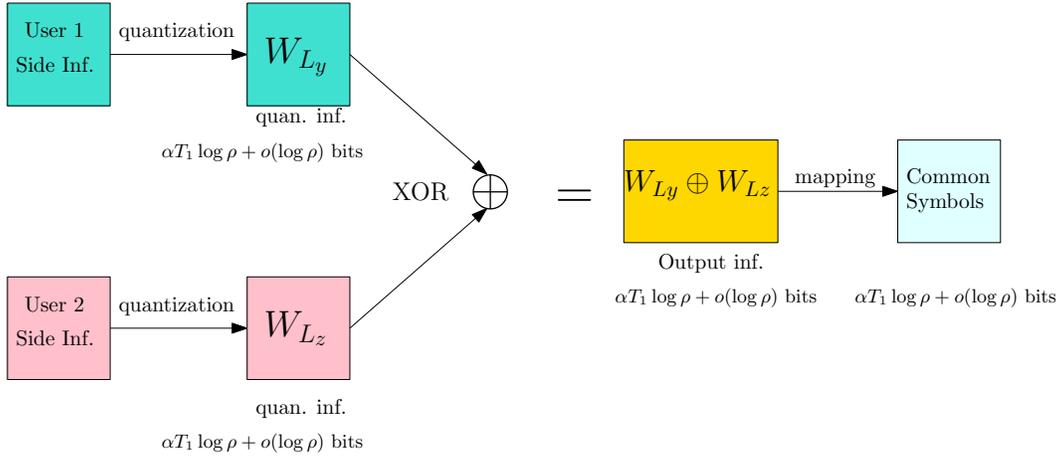}
\caption{Illustration for side information reconstruction and quantization, bitwise XOR operation, and symbol mapping.}
\label{fig:TSM1Mapping}
\end{figure}

\subsubsection{Phase~3}
At the end of the second phase, user~1 knows $\{ y_t=\sqrt{\rho} L_{y} (b_{t,1},  b_{t,2})  +u_t\}_{t=T_1+1}^{T_1+T_2} $, while user~2 knows $\{ z_t=\sqrt{\rho^{\alpha}} L_{z} (a_{t,1},  a_{t,2})  +v_t\}_{t=1}^{T_1} $. At the same time, with the help of delayed CSIT, the transmitter \emph{reconstructs} and \emph{quantizes} the above side information, up to noise level (see Figure~\ref{fig:TSM1Mapping}). Specifically, the transmitter \emph{reconstructs}
\begin{align}   \label{eq:vector1}
    \Bmatrix{ \sqrt{\rho^{\alpha}} L_{z} (a_{1,1},  a_{1,2}) \quad  \sqrt{\rho^{\alpha}} L_{z} (a_{2,1},  a_{2,2})   \  \cdots    \ \sqrt{\rho^{\alpha}} L_{z} (a_{T_1,1},  a_{T_1,2})   }
\end{align}
and \emph{quantizes} the vector using
\begin{align}   \label{eq:qubit1}
   \alpha T_1\log \rho +o(\log \rho)
 \end{align}
quantization bits, allowing for bounded quantization error because $\E|  \sqrt{\rho^{\alpha}} L_{z} (a_{t,1},  a_{t,2})|^2  \doteq  \rho^{\alpha}, \ t=1,2,\cdots,T_1$   (cf. \cite{CT:06}). Similarly the transmitter reconstructs
\begin{align}   \label{eq:vector2}
  \Bmatrix{ \sqrt{\rho} L_{y} (b_{T_1+1 ,1},  b_{T_1+1 ,2})  \quad \sqrt{\rho} L_{y} (b_{T_1+2 ,1},  b_{T_1+2 ,2})    \  \cdots    \  \sqrt{\rho} L_{y} (b_{T_1+T_2 ,1},  b_{T_1+T_2 ,2}) }
\end{align}
and \emph{quantizes} it using
\begin{align}   \label{eq:qubit2}
   T_2\log \rho +o(\log \rho)
   \end{align}
quantization bits, which allows for bounded quantization error since $\E| \sqrt{\rho} L_{y} (b_{t ,1},  b_{t ,2})|^2  \doteq  \rho, \ t=T_1+ 1, \cdots,T_1+T_2$.

Next the transmitter performs the bitwise exclusive-or (XOR) operation on the two sets of quantization bits, i.e., proceeds to bitwise XOR $W_{L_z}$ and $W_{L_y}$ (see Figure~\ref{fig:TSM1Mapping}), where $W_{L_z}$ denotes the vector of $\alpha T_1\log \rho +o(\log \rho)$ quantization bits corresponding to \eqref{eq:vector1}, and where $W_{L_y}$ denotes the vector of (again\footnote{With phase durations designed such that $T_2= T_1 \alpha=T_3\alpha$ (cf. \eqref{eq:phaseduration}), the number of quantization bits in \eqref{eq:qubit1} and \eqref{eq:qubit2} match, and are both equal to $[\alpha T_1\log \rho +o(\log \rho) ]$.} $\alpha T_1\log \rho +o(\log \rho)$) quantization bits corresponding to \eqref{eq:vector2}.

\begin{figure}
\centering
\includegraphics[width=14cm]{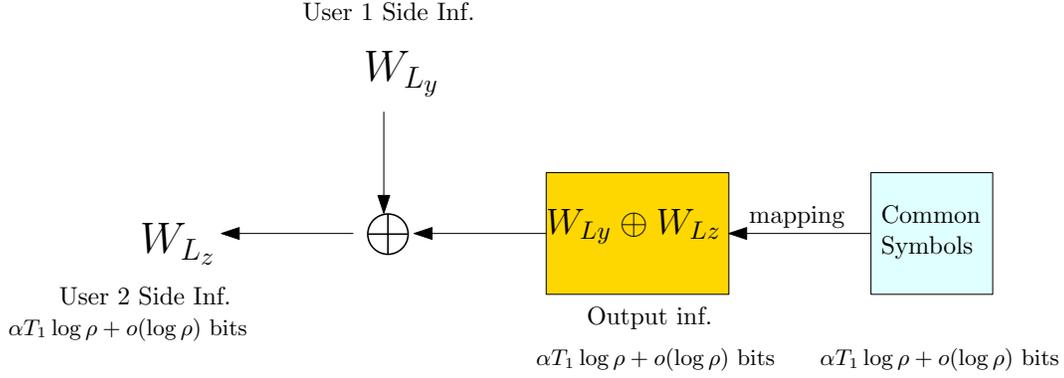}
\caption{Illustration for side information decoding at user~1: Learning user~2's side information from the common information and its side information. }
\label{fig:TSM1User1decodeSI}
\end{figure}

Then the $\alpha T_1\log \rho +o(\log \rho) $ bits in {XOR~($W_{L_z},W_{L_y})$} are mapped into the common symbols $\{ c_t \}$ that will be transmitted in the next phase, in order to eventually allow for recovering the other user's side information (see Figure~\ref{fig:TSM1User1decodeSI}).

As a result, for $t=T_1+T_2+1,\cdots, T_1+T_2+T_3$, the transmitter sends
\begin{align}
\xv_{t} =     \Bmatrix{ c_t  + a_{t,3} \rho^{ - \alpha/2 }    \\   0}
\end{align}
where $c_t$ is a common symbol meant for both users, where $a_{t,3}$ is meant for user 1, where
\begin{equation}
\begin{array}{cc}
P^{(c_t)} \doteq  1,     &  r^{(c_t)} = \alpha \\
P^{(a_{t,3})} \doteq  1, &  r^{(a_{t,3})} = 1-  \alpha
\end{array}
\end{equation}
and where the normalized received signals take the form
\begin{align}
  y_t/ h_{t,1} &=   \underbrace{ \sqrt{\rho}  c_t   }_{\rho}  +\underbrace{ \sqrt{\rho^{1- \alpha}}   a_{t,3}  }_{\rho^{1-\alpha}}     +\underbrace{ u_t /h_{t,1} }_{\rho^{0}}       \label{eq:newMATy103}\\
 z_t/ g_{t,1} &=   \underbrace{ \sqrt{\rho^{\alpha}}  c_t   }_{\rho^{\alpha}}  +\underbrace{ \sqrt{\rho^{0}}   a_{t,3}  }_{\rho^{0}}      +\underbrace{ v_t/ g_{t,1}  }_{\rho^{0}}.   \label{eq:newMATy203}
\end{align}

At this point, for $t=T_1+T_2+1,\cdots, T_1+T_2+T_3$, user~1 can successively decode the common symbol $c_t$ and the private symbol $a_{t,3}$ from $y_t$ (cf. \eqref{eq:newMATy103}), while user~2 can decode the common symbol $c_t$ from $z_t$ by treating the other signals as noise (cf. \eqref{eq:newMATy203}).

Upon decoding $\{c_t\}_{t=T_1+T_2+1}^{T_1+T_2+T_3}$, user~1 can recover {XOR~($W_{L_z}, W_{L_y}$)}, and can thus sufficiently-well recover $W_{L_z}$ using its own side information $ \{ y_t\}_{t=T_1+1}^{T_1+T_2} $, thus recovering $\{ \sqrt{\rho^{\alpha}} L_{z} (a_{t,1},  a_{t,2}) \}_{t=1}^{T_1} $ up to noise level.
This in turn allows user~1 to obtain the following `MIMO observations' for $t=1,2,\cdots,T_1$
\begin{align} \label{eq:MIMOnewMATdCSIT}
\begin{bmatrix} y_t
          \\     \sqrt{\rho^{\alpha}} L_{z} (a_{t,1},  a_{t,2})  + \tilde{\iota}_{z,t}  \end{bmatrix}   =   \Bmatrix{ \sqrt{\rho}\hv^\T_t  \\  \sqrt{\rho^{\alpha}} \gv^\T_t  }   \Bmatrix{ a_{t,1} \\  a_{t,2} }
 + {\begin{bmatrix}  u_t \\
              \tilde{\iota}_{z,t} \end{bmatrix}}
\end{align}
and to MIMO decode $a_{t,1}, a_{t,2}$ at the declared rates (cf. \eqref{eq:TSMDD1aPowerrate1}). In the above we used $\tilde{\iota}_{z,t}$ to denote the aforementioned quantization and reconstruction noise, which - given the number of quantization bits - can be seen to have bounded power.

Similarly, upon decoding $\{c_t\}_{t=T_1+T_2+1}^{T_1+T_2+T_3}$, user~2 uses $\{ z_t\}_{t=1}^{T_1} $ to recover $\{ \sqrt{\rho} L_{y} (b_{t,1},  b_{t,2}) \}_{t=T_1+1}^{T_1+T_2} $ sufficiently well, and to allow for a MIMO observation
\begin{align} \label{eq:MIMOnewMATdCSIT2}
\begin{bmatrix} z_t
          \\    \sqrt{\rho} L_{y} (b_{t,1},  b_{t,2})    +  \tilde{\iota}_{y,t}    \end{bmatrix}   =  \Bmatrix{  \sqrt{\rho^{\alpha}} \gv^\T_t \\  \sqrt{\rho} \hv^\T_t  }   \Bmatrix{ b_{t,1} \\  b_{t,2} }
 + {\begin{bmatrix}  v_t \\
              \tilde{\iota}_{y,t}   \end{bmatrix}}
\end{align}
which results in the subsequent decoding of $b_{t,1}, b_{t,2}$ ($t=T_1+1,\cdots,T_1+T_2$) at the declared rates \eqref{eq:TSMDD1aPowerrate2}. In the above, we used $\tilde{\iota}_{y,t}$ to denote the previous quantization and reconstruction noise, which can be shown to have bounded power.

As a result, summing up the number of information bits, allows us to conclude that the proposed scheme achieves a sum GDoF
\begin{align*}
d_{\sum}&=\frac{T_1 (1+ \alpha ) +  T_2 (1+\alpha ) + T_3 (1 - \alpha)}{T_1+T_2+T_3} \\
&=\frac{2+  \alpha+ \alpha^2}{2+ \alpha }   \\
&=1+  \frac{ \alpha^2}{2+ \alpha }.
\end{align*}

\subsection{TSM scheme for $\lambda_{D, D}^{ 1,\alpha }=\lambda_{D, D}^{ \alpha,1 }=1/2$, achieving the optimal sum GDoF $(1+\alpha/3)$   \label{sec:DD1a-DDa1}}

We now transition to an alternating topology.

The scheme can be described as having three channel uses, $t=1,2,3$. We will first, without loss of generality, describe the scheme for the setting where, for $t=1,3$, the feedback-and-topology state is $(I_1,I_2,A_1,A_2)=(D, D, 1, \alpha )$, and for $t=2$ the state is $(I_1,I_2,A_1,A_2)=(D, D, \alpha,1 )$. The scheme can be slightly modified for the case where $(I_1,I_2,A_1,A_2)=\underbrace{(D, D, 1, \alpha )}_{t=1},  \underbrace{(D, D,  \alpha,1)}_{t=2},  \underbrace{(D, D,  \alpha, 1)}_{t=3}$. In both cases, the scheme can achieve the optimal sum GDoF $(1+\alpha/3)$.

\begin{figure}
\centering
\includegraphics[width=12cm]{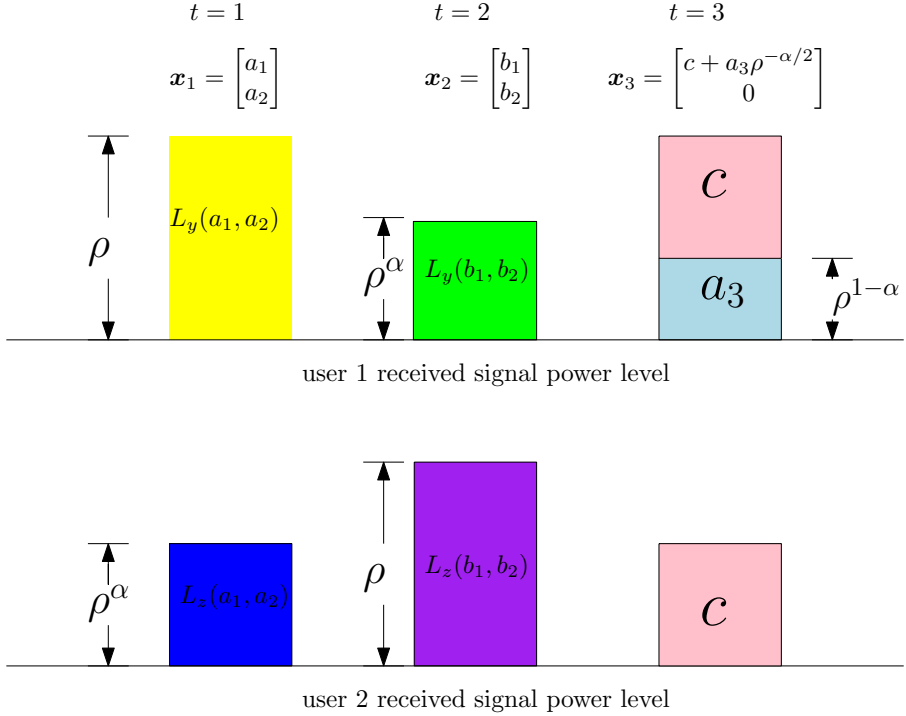}
\caption{ Received signal power level illustration for the TSM scheme, for the setting where $\lambda_{D, D}^{ 1,\alpha }=\lambda_{D, D}^{ \alpha,1 }=1/2$. }
\label{fig:TSM2Top}
\end{figure}

\subsubsection{Phase~1} At $t=1$ $((I_1,I_2,A_1,A_2)= (D, D, 1, \alpha )$, link 1 is strong$)$ the transmitter sends (see Figure~\ref{fig:TSM2Top})
\begin{align}
\xv_{1} =   \Bmatrix{ a_1 \\  a_2 }
\end{align}
where  $a_1$ and $a_2$ are unit-power symbols meant for user 1, with
\begin{equation}  \label{eq:DD1aa1PowerP1}
\begin{array}{cc}
  r^{(a_1)} = 1, &  r^{(a_2)} = \alpha
\end{array}
\end{equation}
resulting in received signals of the form
\begin{align}
  y_1&=   \underbrace{\sqrt{\rho}    \hv^\T_1 \Bmatrix{ a_1 \\  a_2 }   }_{\rho} +u_1 \label{eq:sch22y1}\\
  z_1	&= \underbrace{ \sqrt{\rho^{\alpha}}\gv^\T_1 \Bmatrix{ a_1 \\  a_2 }   }_{\rho^{\alpha}}    +v_1 \label{eq:sch22y2}
\end{align}
where we note that the unintended interfering signal is attenuated due to the weak link.
\subsubsection{Phase~2} At time $t=2$ $((I_1,I_2,A_1,A_2)= (D, D, \alpha, 1)$, link 1 is weak$)$ the transmitter sends
\begin{align}
\xv_{2} =     \Bmatrix{ b_1 \\  b_2 }
\end{align}
where $b_1,b_2$ are unit-power symbols meant for user 2, with
\begin{equation}  \label{eq:DD1aa1PowerP2}
\begin{array}{cc}
  r^{(b_1)} = 1, &  r^{(b_2)} = \alpha
\end{array}
\end{equation}
resulting in received signals of the form
\begin{align}
  y_2&=   \underbrace{ \sqrt{\rho^{\alpha}}\hv^\T_2 \Bmatrix{ b_1 \\  b_2 }   }_{\rho^{\alpha}}   +u_2 \label{eq:sch22y102}\\
  z_2	&=  \underbrace{\sqrt{\rho}    \gv^\T_2 \Bmatrix{ b_1 \\  b_2 }   }_{\rho}  +v_2 \label{eq:sch22y202}
\end{align}
where again the unintended interfering signal is attenuated due to the weak link.

\subsubsection{Phase~3}

At this point the transmitter - using delayed CSIT - knows $\gv_1 $ and $\hv_2$. It then proceeds to reconstruct $(z_1 - v_1)$ and $(y_2 -u_2)$, and to quantize the sum
\begin{align}   \label{eq:iotadef}
\iota \defeq (z_1 - v_1) +(y_2 -u_2)
\end{align}
using $\alpha\log \rho + o(\log \rho)$ quantization  bits, in order to get the quantized version $\bar{\iota}$. Given the number of quantization bits, and given that $\E|\iota|^2 \doteq \rho^{\alpha}$, the quantization error \[\tilde{\iota} =\iota - \bar{\iota} \] is bounded and does not scale with $\rho$ (cf. \cite{CT:06}). The above quantized information is then mapped into a \emph{common} symbol $c$.

At time $t=3$, with state $(I_1,I_2,A_1,A_2)= (D, D, 1,  \alpha)$ (link 2 is weak), the transmitter sends
\begin{align}   \label{eq:sch2873t3}
\xv_{3} =     \Bmatrix{ c  + a_3 \rho^{ - \alpha/2 }    \\   0}
\end{align}
where $c$ is the aforementioned common symbol meant for both users, where $a_3$ is a symbol meant for user 1, where
\begin{equation}
\begin{array}{cc}
P^{(c)} \doteq  1,     &  r^{(c)} = \alpha \\
P^{(a_3)} \doteq  1, &  r^{(a_3)} = 1-  \alpha
\end{array}
\end{equation}
and where the (normalized) received signals (in their noiseless form) are
\begin{align}
  y_3/ h_{3,1} &=   \sqrt{\rho}  c    + \sqrt{\rho^{1- \alpha}}   a_3   \label{eq:sch22y103}\\
 z_3/ g_{3,1} &=   \sqrt{\rho^{\alpha}}  c     +   \sqrt{\rho^{0}}   a_3.  \label{eq:sch22y203}
\end{align}

Now we see from \eqref{eq:sch22y103},\eqref{eq:sch22y203} that $c$ can be decoded by both users. Similarly we can readily see that $a_3$ can be decoded by user~1.

At this point, knowing $c$ allows both users to recover $\bar{\iota}$ (cf. \eqref{eq:iotadef}), and to then decode the private symbols. Specifically, user~1 obtains a MIMO observation
\begin{align} \label{eq:MIMOdCSIT}
\begin{bmatrix} y_1
          \\      \bar{\iota}  -   y_2\end{bmatrix}   = \Bmatrix{ \sqrt{\rho} \hv^\T_1 \\ \sqrt{\rho^{\alpha}}\gv^\T_1  }   \Bmatrix{ a_1 \\  a_2 }
 + {\begin{bmatrix}  u_1 \\
              - u_2- \tilde{\iota} \end{bmatrix}}
\end{align}
which allows for decoding of $a_1, a_2$ at the declared rates (cf. \eqref{eq:DD1aa1PowerP1}).
Similarly, user~2 obtains another MIMO observation
\begin{align} \label{eq:MIMOdCSIT2}
\begin{bmatrix} z_2
          \\      \bar{\iota} -   z_1\end{bmatrix}   = \Bmatrix{ \sqrt{\rho} \gv^\T_2 \\ \sqrt{\rho^{\alpha}}\hv^\T_2  }   \Bmatrix{ b_1 \\  b_2 }
 + {\begin{bmatrix}  v_2 \\
              - v_1- \tilde{\iota} \end{bmatrix}}
\end{align}
and can decode $b_1, b_2$ at the declared rates (cf. \eqref{eq:DD1aa1PowerP2}).
Summing up the information bits concludes that the scheme achieves the optimal sum GDoF $d_{\sum}=\frac{1 + \alpha +1  + \alpha  + (1- \alpha)}{3}=1+ \frac{\alpha}{3}$ (also see Figure~\ref{fig:TSM2Top}).

\begin{remark}
As stated above, when $(I_1,I_2,A_1,A_2)= (D, D, 1,\alpha ),  (D, D, \alpha,1),  (D, D, \alpha, 1) $ for $t=1,2,3$ respectively, we can slightly modify the scheme such that at $t=3$, instead of sending the private symbol $a_3$ for the first user (see \eqref{eq:sch2873t3}), to instead send a private symbol $b_3$ for the second user (i.e., again to the stronger user). Following the same steps, one can easily show that the sum GDoF $d_{\sum}=1 + \alpha/3$ is again achievable.
\end{remark}

\begin{remark}
It is interesting to note that the proposed scheme needs delayed CSIT for only a fraction of the channels (the channels with weak channel gain in phase~1 and phase~2), and in essence only needs $\lambda_{N,D}^{1,\alpha}=\lambda_{D,N}^{\alpha,1}=\lambda_{N,N}^{1,\alpha}= 1/3$, or $\lambda_{N,D}^{1,\alpha}=\lambda_{D,N}^{\alpha,1}=\lambda_{N,N}^{\alpha,1}= 1/3$, or $\lambda_{N,D}^{1,\alpha}=\lambda_{D,N}^{\alpha,1}=\frac{1}{2}\lambda_{N,N}^{1,\alpha}=\frac{1}{2}\lambda_{N,N}^{\alpha,1}= 1/3$, to achieve the same optimal sum GDoF.
\end{remark}

\subsection{TSM schemes for $\lambda_{P,N}=\lambda_{N,P} = 1/2$ and for any $\lambda_{1,\alpha}+\lambda_{\alpha,1}=1$; achieving the optimal sum GDoF $ 1+  \frac{ \alpha}{2}$\label{sec:PNNP-any}}

We will now show that the optimal sum GDoF $ (1+  \frac{ \alpha}{2})$ is achievable for any topology $\lambda_{1,\alpha}+\lambda_{\alpha,1}=1$ using $\lambda_{P,N}=\lambda_{N,P} = 1/2$ and a sequence of TSM schemes proposed for the different settings of
\[\lambda_{P,N}^{1, \alpha}  = \lambda_{N,P}^{1, \alpha} =1/2; \  \ \lambda_{P,N}^{\alpha,1}  = \lambda_{N,P}^{ \alpha,1} =1/2; \  \ \lambda_{P,N}^{1, \alpha}  = \lambda_{N,P}^{\alpha,1} =1/2; \  \ \lambda_{P,N}^{\alpha,1}  = \lambda_{N,P}^{1, \alpha} =1/2\] respectively.
Each scheme achieves  the optimal sum GDoF $ (1+  \frac{ \alpha}{2})$, and each scheme is designed to have only two channel uses, during which the two users take turn to feed back current CSIT (only one user feeds back at a time).
The general result is proven by properly concatenating the proposed schemes for the different cases.

\subsubsection{TSM scheme for $\lambda_{P,N}^{1, \alpha}  = \lambda_{N,P}^{1, \alpha} =1/2$ \label{sec:PN1a-NP1a}}

\begin{figure}
\centering
\includegraphics[width=10cm]{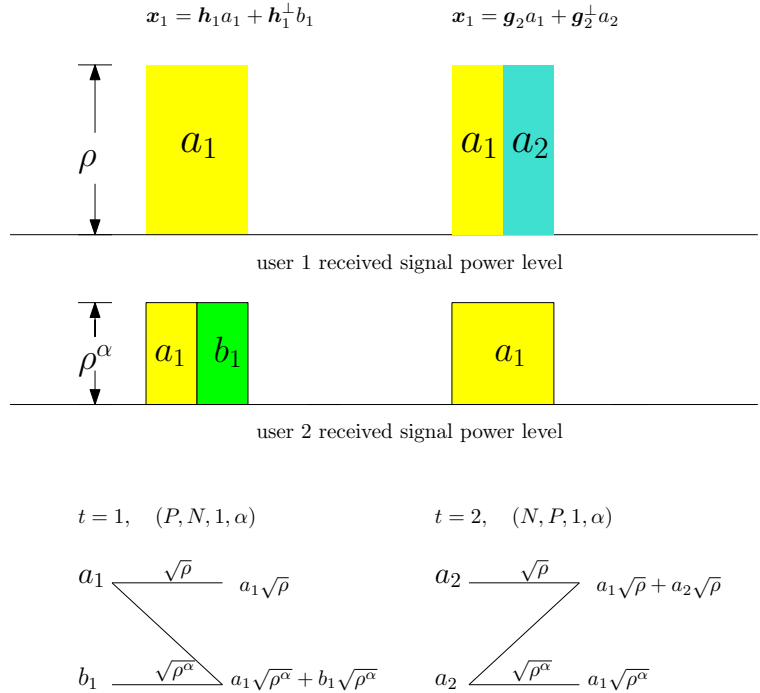}
\caption{Illustration of TSM coding and of received signal power levels, for $\lambda_{P,N}^{1, \alpha}  = \lambda_{N,P}^{1, \alpha} =1/2$. }
\label{fig:TSMPNNPX31}
\end{figure}

Without loss of generality, we focus on the specific sub-case where $(I_1,I_2,A_1,A_2)= (P, N,  1 ,  \alpha)$ for $t=1$, and $(I_1,I_2,A_1,A_2) = (N, P,  1 ,  \alpha)$ for $t=2$.

At $t=1$ the transmitter knows $\hv_1$ (current CSIT), and sends (see Figure~\ref{fig:TSMPNNPX31})
\begin{align} \label{eq:signalX101}
\xv_{1} =     \hv_1 a_1 +  \hv_1^\bot  b_1
\end{align}
where $a_1$ and $b_1$ are intended for user 1 and user~2 respectively, and where
\begin{equation}\label{eq:RPX1101}
\begin{array}{cc}
P^{(a_1)} \doteq  1, &  r^{(a_1)} = 1 \\
P^{(b_1)} \doteq  1, &  r^{(b_1)} = \alpha.
\end{array}
\end{equation}
Then the received signals (in their noiseless form) are
\begin{align}
  y_1&=   \underbrace{\sqrt{\rho}    \hv^\T_1 \hv_1 a_1}_{\rho} \label{eq:sch1y1}\\
  z_1	&= \underbrace{ \sqrt{\rho^{\alpha}}\gv^\T_1 \hv_1 a_1}_{\rho^{\alpha}}  +\underbrace{ \sqrt{\rho^{\alpha} }\gv^\T_1 \hv_1^\bot b_1}_{\rho^{\alpha}}. \label{eq:sch1y2}
\end{align}

At $t=2$ $((I_1,I_2,A_1,A_2) = (N, P,  1 ,  \alpha))$, the transmitter knows $\gv_2$ (current CSIT) and sends
\begin{align} \label{eq:signalX102}
\xv_{2} =     \gv_2 a_1 +  \gv_2^\bot  a_2
\end{align}
where $a_2$ is intended for user 1, and where
\begin{equation}\label{eq:RPX1102}
\begin{array}{cc}
P^{(a_2)} \doteq  1, &  r^{(a_2)} = 1.
\end{array}
\end{equation}
Then the received signals (in their noiseless form) are as follows
\begin{align}
  y_2&=   \underbrace{\sqrt{\rho}    \hv^\T_2 \gv_2 a_1}_{\rho} +\underbrace{\sqrt{\rho}    \hv^\T_2 \gv^\bot_2 a_2}_{\rho} \label{eq:sch1y1323}\\
  z_2	&= \underbrace{ \sqrt{\rho^{\alpha}}\gv^\T_2 \gv_2 a_1}_{\rho^{\alpha}}.      \label{eq:sch1y23123}
\end{align}
At this point, we can see that user~1 can MIMO decode $a_1, a_2$ based on \eqref{eq:sch1y1}, \eqref{eq:sch1y1323}, while user~2 can recover $b_1$ by employing interference cancelation based on \eqref{eq:sch1y2},  \eqref{eq:sch1y23123}.  This gives a sum DoF of $1 + \alpha/2$.

\begin{remark}
We can now readily see that for the setting where $(I_1,I_2,A_1,A_2)= \overbrace{(N, P, 1 ,  \alpha)}^{t=1},  \overbrace{(P, N, 1 ,  \alpha)}^{t=2}$, we can easily modify the above scheme to achieve the same performance, just by reordering the transmissions such that $\xv_{1} =     \gv_1 a_1 +  \gv_1^\bot  a_2 $ and $\xv_{2} =     \hv_2 a_1 +  \hv_2^\bot  b_1$.

Similarly when $\lambda_{P,N}^{ \alpha,1}  = \lambda_{N,P}^{\alpha,1} =1/2$, we can take the above scheme (of Section~\ref{sec:PN1a-NP1a}), and simply interchange the roles of the users, to again achieve the optimal sum GDoF $1 + \alpha/2$.
\end{remark}

\subsubsection{TSM scheme for $\lambda_{P,N}^{1, \alpha}  = \lambda_{N,P}^{\alpha,1} =1/2$ \label{sec:PN1a-NPa1}}

\begin{figure}
\centering
\includegraphics[width=12cm]{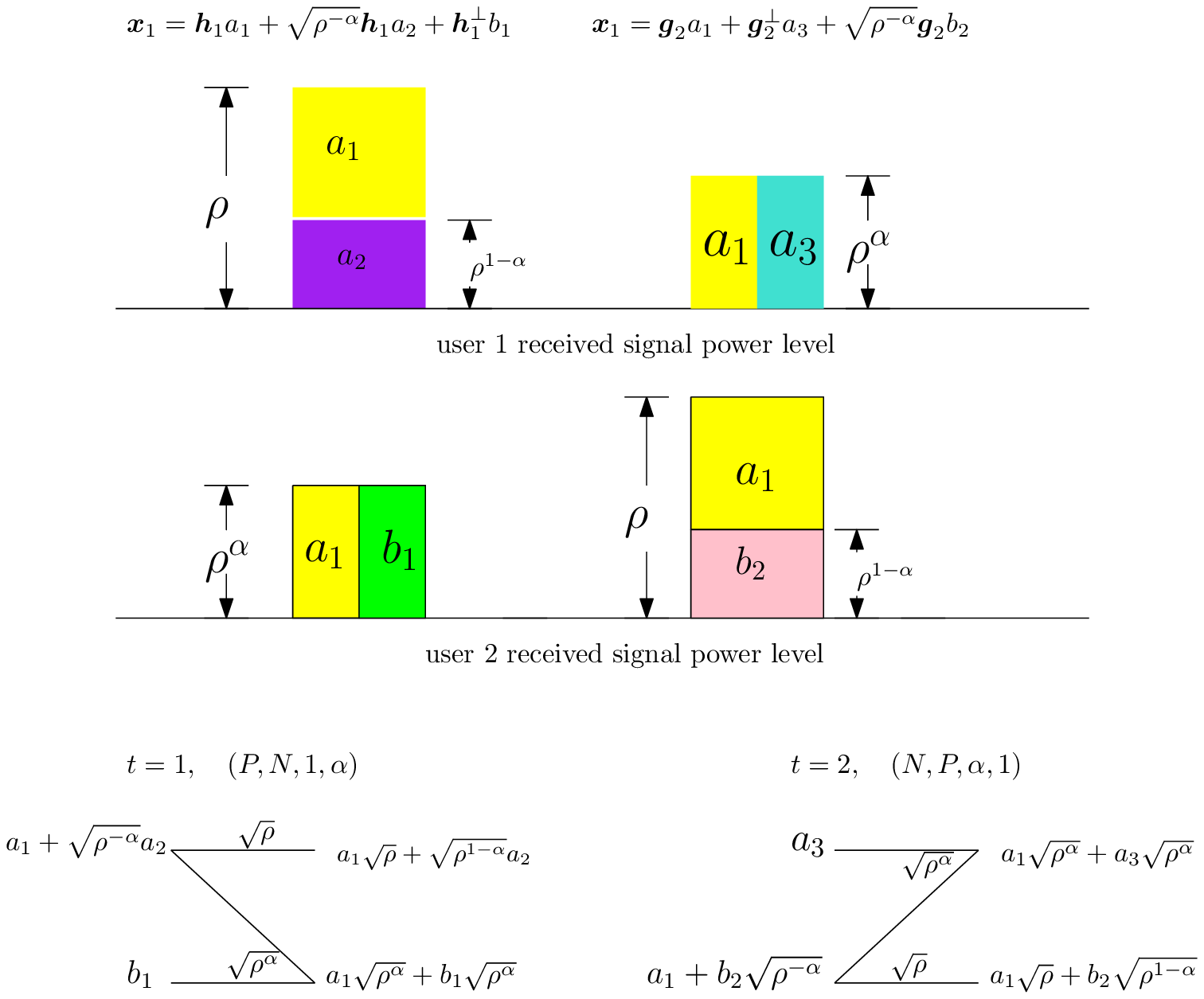}
\caption{Illustration of coding and received signal power levels for $\lambda_{P,N}^{1, \alpha}  = \lambda_{N,P}^{\alpha,1} =1/2$.}
\label{fig:TSMPNNPX32}
\end{figure}

We focus on the case where we first have $(I_1,I_2,A_1,A_2)= (P, N,  1 ,  \alpha)$ (at $t=1$), followed by $(I_1,I_2,A_1,A_2)= (N, P, \alpha,1)$ ($t=2$).

At $t=1$, the transmitter knows $\hv_1$, and sends (see Figure~\ref{fig:TSMPNNPX32})
\begin{align}
\xv_{1} =     \hv_1 a_1 + \sqrt{\rho^{-\alpha}} \hv_1  a_2   +  \hv_1^\bot  b_1
\end{align}
where $a_1, a_2$ are the unit-power symbols intended for user~1, $b_1$ is the unit-power symbol intended for user 2, where
\begin{equation}\label{eq:PNNP1aa1RP1}
\begin{array}{ccc}
  r^{(a_1)} = \alpha, &  r^{(a_2)} = 1-\alpha, &  r^{(b_1)} = \alpha
\end{array}
\end{equation}
and where the received signals, in their noiseless form, are
\begin{align}
  y_1&=   \underbrace{\sqrt{\rho}    \hv^\T_1 \hv_1 a_1}_{\rho}  + \underbrace{\sqrt{\rho^{1-\alpha}}    \hv^\T_1 \hv_1 a_2}_{\rho^{1-\alpha}}  \label{eq:PNNP1aa1y1}\\
  z_1	&= \underbrace{ \sqrt{\rho^{\alpha}}\gv^\T_1 \hv_1 a_1}_{\rho^{\alpha}}  + \underbrace{\sqrt{\rho^{0}}    \gv^\T_1 \hv_1 a_2}_{\rho^{0}}    +\underbrace{ \sqrt{\rho^{\alpha} }\gv^\T_1 \hv_1^\bot b_1}_{\rho^{\alpha}}.    \label{eq:PNNP1aa1y2}
\end{align}

At $t=2$ $((I_1,I_2,A_1,A_2)= (N, P, \alpha,1))$ the transmitter knows $\gv_2$ (user~1 is weak), and sends
\begin{align}
\xv_{2} =     \gv_2 a_1 +  \gv_2^\bot  a_3  +  \sqrt{\rho^{-\alpha}}  \gv_2 b_2
\end{align}
where $a_3, b_2$ are the unit-power symbols intended for user~1 and user~2 respectively, where
\begin{equation}\label{eq:PNNP1aa1RP2}
\begin{array}{cc}
  r^{(a_3)} = \alpha, &  r^{(b_2)} = 1-\alpha
\end{array}
\end{equation}
and where the received signals, in their noiseless form, are
\begin{align}
  y_2&=   \underbrace{\sqrt{\rho^{\alpha}}    \hv^\T_2 \gv_2 a_1}_{\rho^{\alpha}} +\underbrace{\sqrt{\rho^{\alpha}}    \hv^\T_2 \gv^\bot_2 a_3}_{\rho^{\alpha}}   +\underbrace{\sqrt{\rho^{0}}    \hv^\T_2 \gv^\bot_2 b_2}_{\rho^{0}}  \label{eq:PNNP1aa1y11}\\
  z_2	&= \underbrace{ \sqrt{\rho}\gv^\T_2 \gv_2 a_1}_{\rho}   + \underbrace{ \sqrt{\rho^{1-\alpha}}\gv^\T_2 \gv_2 b_2}_{\rho^{1-\alpha}}.    \label{eq:PNNP1aa1y22}
\end{align}

At this point, it is easy to see that user~1 can recover $a_1, a_2, a_3$ by MIMO decoding based on \eqref{eq:PNNP1aa1y1} and \eqref{eq:PNNP1aa1y11}, while user~2 can recover $b_1, b_2$ by employing interference cancelation based on \eqref{eq:PNNP1aa1y2} and \eqref{eq:PNNP1aa1y22} (see also Figure~\ref{fig:TSMPNNPX32}).  This provides for $d_{\sum}=1 + \alpha/2$.

\paragraph{Modifying the scheme for the setting where $(I_1,I_2,A_1,A_2)$ is $(N, P, \alpha, 1 )$ or $(P, N, 1 ,  \alpha) $}
Similarly for the setting where $(I_1,I_2,A_1,A_2)$ is $(N, P, \alpha, 1 )$ or $(P, N, 1 ,  \alpha) $, we can modify the previous scheme --- to achieve the same optimal sum DoF --- by interchanging the transmissions of the first and second channel uses, i.e., of $t=1, 2$.

\paragraph{Modifying the scheme for the setting where $\lambda_{P,N}^{\alpha,1}  = \lambda_{N,P}^{1,\alpha} =1/2$}
Furthermore when $\lambda_{P,N}^{\alpha,1}  = \lambda_{N,P}^{1,\alpha} =1/2$, we can simply interchange the roles of users in the previous scheme, to again achieve the same optimal sum GDoF.

\paragraph{Spanning the entire setting $\lambda_{1, \alpha}  + \lambda_{\alpha,1} =1$, $\lambda_{P,N}=\lambda_{N,P}$}
Finally, by using $\lambda_{P,N}=\lambda_{N,P}$ and by properly concatenating the above scheme variants, gives the optimal performance $d_{\sum}=1 + \alpha/2$, for the entire range $\lambda_{1, \alpha}  + \lambda_{\alpha,1} =1$.

\subsection{Original MAT scheme in the fixed topological setting ($\lambda_{1,\alpha }=1$)\label{sec:MATorig-DD1a}}

\begin{figure}
\centering
\includegraphics[width=12cm]{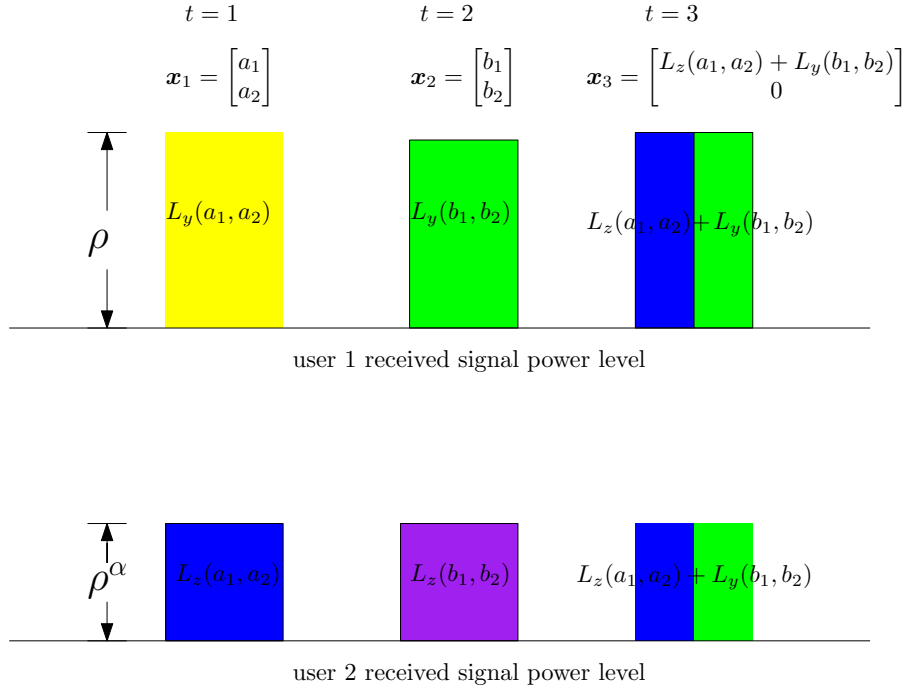}
\caption{Illustration of received power level for the original MAT scheme in the fixed topology setting $\lambda_{1,\alpha}=1$.}
\label{fig:MATsignal}
\end{figure}

We recall that the original MAT scheme in \cite{MAT:11c} consists of three phases, each of duration one.
At time $t=1,2$, the transmitter sends
\[ \xv_{1} =   \Bmatrix{ a_1 \\  a_2 }, \quad \quad \xv_{2} =     \Bmatrix{ b_1 \\  b_2 } \]
where $a_1,a_2$ are for user 1, $b_1,b_2$ for user 2, and where the received signals, in their noiseless form, are now (in the current, topologically sensitive setting)
\begin{align}
  y_1&=   \sqrt{\rho}    \hv^\T_1 \Bmatrix{ a_1 \\  a_2 }    \quad \quad \quad  \quad \quad \quad  \quad \quad \quad           z_1	=  \sqrt{\rho^{\alpha}}\gv^\T_1 \Bmatrix{ a_1 \\  a_2 }     \defeq    \sqrt{\rho^{\alpha}}  L_{z} (a_1,  a_2)     \label{eq:MAT001y2}   \\
y_2&=    \sqrt{\rho}\hv^\T_2 \Bmatrix{ b_1 \\  b_2 }   \defeq  \sqrt{\rho}   L_{y} (b_1,  b_2)    \quad \quad \quad
  z_2	=  \sqrt{\rho^{\alpha}}    \gv^\T_2 \Bmatrix{ b_1 \\  b_2 }.   \label{eq:MAT001y102}
\end{align}
At $t=3$, the transmitter knows $\gv_1 $ and $\hv_2$ (delayed CSIT), reconstructs  $L_{z} (a_1,  a_2)  ,L_{y} (b_1,  b_2)  $  (cf. \eqref{eq:MAT001y2}, \eqref{eq:MAT001y102}), and sends  \[\xv_{3} =     \Bmatrix{ L_{z} (a_1,  a_2) + L_{y} (b_1,  b_2)  \\   0} \]
with normalized/processed received signals which, in their noiseless form, are
\begin{align}
  y_3/ h_{3,1} &=   \sqrt{\rho}  L_{z} (a_1,  a_2)   +  \sqrt{\rho}  L_{y} (b_1,  b_2)     \\
 z_3/ g_{3,1} &=    \sqrt{\rho^{\alpha}}  L_{z} (a_1,  a_2) +  \sqrt{\rho^{\alpha}}  L_{y} (b_1,  b_2).
\end{align}
At this point, we recall from~\cite{MAT:11c} that user~1 combines the above with $y_1,y_2, y_3$, to design a MIMO system
\begin{align} \label{eq:MIMOMATdCSIT}
\begin{bmatrix} y_1
          \\      y_3/ h_{3,1}    -   y_2\end{bmatrix}   = \sqrt{\rho}  \Bmatrix{ \hv^\T_1  \\  \gv^\T_1  }   \Bmatrix{ a_1 \\  a_2 }
 + {\begin{bmatrix}  u_1 \\
              u_3/h_{3,1} - u_2 \end{bmatrix}}
\end{align}
and to MIMO decode $a_1, a_2$, which carry a total of $[2\log \rho + o(\log \rho)  ]$ bits.
Similarly, user~2 is presented with another MIMO system
\begin{align} \label{eq:MIMOMATdCSIT2}
\begin{bmatrix} z_2
          \\      z_3/ g_{3,1}  -   z_1\end{bmatrix}   =\sqrt{\rho^{\alpha}}  \Bmatrix{  \gv^\T_2 \\ \hv^\T_2  }   \Bmatrix{ b_1 \\  b_2 }
 + {\begin{bmatrix}  v_2 \\
              v_3/g_{3,1}  - v_1 \end{bmatrix}}
\end{align}
of less power, from which it can MIMO decode $b_1, b_2$, which though now carry a total of $2\alpha\log \rho + o(\log \rho)$ bits.
As a result, the original MAT scheme achieves a sum GDoF $d_{\sum}=\frac{2(1 + \alpha)}{3}$.


\end{document}